\documentclass[aps,prl,twocolumn,showpacs,superscriptaddress]{revtex4-1}

\usepackage{amsmath} % AMS Math Package
\usepackage{amsthm} % Theorem Formatting
\usepackage{amssymb}	% Math symbols such as \mathbb
\usepackage{graphicx} % Allows for eps images
\usepackage[table]{xcolor}
\usepackage{comment}
\usepackage{filecontents} 
\usepackage{stackengine}

\stackMath

\newcommand{\ket}[1]{\left| #1 \right>} % for Dirac bras
\newcommand{\bra}[1]{\left< #1 \right|} % for Dirac kets
\newcommand{\trace}[1]{\textrm{tr} \left[ #1 \right]} % Trace

\newtheorem{theorem}{Theorem}

\begin{document}
\title{Dynamical engineering of interactions in qudit ensembles} 

\author{Soonwon Choi}
\affiliation{Department of Physics, Harvard University, Cambridge, Massachusetts 02138, USA}

\author{Norman Y. Yao}
\affiliation{Department of Physics, University of California Berkeley, Berkeley, California 94720, USA}

\author{Mikhail D. Lukin}
\affiliation{Department of Physics, Harvard University, Cambridge, Massachusetts 02138, USA}

\date{\today}

\begin{abstract}
We propose and analyze a method to engineer effective interactions in an ensemble of $d$-level systems (qudits) driven by global control fields.
In particular, we present (i) a necessary and sufficient condition under which a given interaction can be turned off (decoupled), (ii) the existence of a universal sequence that decouples any (cancellable) interaction, and (iii) an efficient algorithm to engineer a target Hamiltonian from an initial Hamiltonian (if possible).
As examples, we provide a 6-pulse sequence that decouples effective spin-1 dipolar interactions and demonstrate that a spin-1 Ising chain can be engineered to study transitions among three distinct symmetry protected topological phases.
\end{abstract}

\maketitle

The controlled manipulation of quantum systems with pulsed coherent fields 
is important in nearly all branches of quantum science. 
In its simplest form, such manipulation involves the time-dependent modulation of a quantum system, with the aim of steering the system's dynamics.
The techniques associated with dynamical coherent control have a long and storied history, originating in nuclear magnetic resonance (NMR), where periodic sequences of instantaneous control pulses enable the isolation of nuclear spins from unwanted external noise sources~\cite{Hahn_echo:1950ge}.
Over the past few decades, advanced techniques have been developed with goals ranging from frequency-selective decoupling to higher-order error suppression, and applications ranging from metrology to information processing \cite{Uhrig:2007ek,deLange:2010ga,QDDproofLidar:2011aa,UDDproofJiang:2011aa,NV_memory:2012Maurer,PazSilva:2013hx,entangledQubitClock2014,Lovchinsky:2016dq,Lovchinsky:2017gh}.

Periodic control pulses can also be used to engineer many-body interactions. In particular, they can enable the realization of 
driven (Floquet) system that exhibit phenomena richer than the original system without dynamical control~\cite{Lindner:2011ip,FloquetMajoranaJiang:2011aa,Iadecola:2015dg,Khemani:2015gd,Else:2016gf,vonKeyserlingk:2016ev,Yao_dtc:2016wp}. 
This approach falls under the moniker of average Hamiltonian theory~\cite{Haeberlen:1968zz}, a term prevalent in the context of solid-state NMR, where sequences of spin-rotations are used to modify the intrinsic interactions between magnetic dipoles \cite{Waugh:1968im,Haeberlen:1968zz}.
A particularly powerful example is the celebrated WAHUHA pulse sequence~\cite{Waugh:1968im} which cancels the dipole-dipole interaction between spin-1/2 particles and has been extensively utilized in systems ranging from solid-state spin defects to ultracold polar molecules~\cite{NV_memory:2012Maurer,Yan:2013fn}. 
While the majority of existing pulse sequences are designed to engineer Hamiltonians constructed from spin-1/2 or qubit-like systems~\cite{Brinkmann:2004fk,Ajoy:2013hf,Frydrych:2014il,Hayes:2014vj}, recent experimental progress has opened the door to the  manipulation of many-body  \emph{qudit} systems, whose basic degrees of freedom possess $d$ internal states.
Indeed, in platforms ranging from trapped ions and Rydberg atoms to superconducting qubits and solid-state spin defects, coherent interactions among multiple qudits have already been observed~\cite{Yan:2013fn,Senko:2015hf,Choi:2016wn}.
This enables the study of quantum many-body qudit systems that can exhibit phenomena  qualitatively distinct from their spin-1/2 counterparts, such as  generalized Potts model and parafermionic topological phases \cite{Potts:1951be,Huse:1981gg,Haldane:1983ru,Fendley:2012hw}.
Generalizing Hamiltonian engineering methods to qudit systems may enable exploration of such unique phenomena in dynamical systems with important potential applications in areas such as  quantum simulations.

In this Letter, we report two advances toward this goal.
First, we present a generalization of the WAHUHA pulse sequence for an arbitrary qudit system. 
In particular, we derive a necessary and sufficient condition that diagnoses when generic interactions in a qudit system can be cancelled. 
Moreover, we prove the existence of a universal pulse sequence that decouples any cancelable interaction. 
As a specific example, we present a novel pulse sequence that decouples spin-1 dipolar interactions.
Second, we present an algorithm that uniquely determines when a given initial qudit Hamiltonian $H_0$ can be mapped to a desired final Hamiltonian $H_f$, using  a predetermined set of global pulses. 
In this context, we demonstrate that a spin-1 Ising chain can be directly mapped to a family of Hamiltonians whose ground states include a variety of symmetry protected topological (SPT) phases.
In both cases, we consider an ensemble of $d$-level systems with generic pairwise interactions and assume that only \emph{global} $SU(d)$ manipulations are available.
We note that in the case where qudits can be \emph{independently} addressed and controlled,  arbitrary modifications of the underlying interactions are possible~\cite{Rotteler:2006km,PhysRevA.64.052301,PhysRevA.66.022317,Frydrych:2014il,Hayes:2014vj}; however, such precise individual controls are typically challenging to implement in strongly interacting many-body systems. 
We consider an $N$ qudit system with Hamiltonian,
\begin{align}
\label{eqn:generic_hamiltonian}
    H = \sum_{ij} J_{ij} h_{ij},
\end{align}
where $h_{ij}$ represents a  homogeneous two-qudit interaction between $i$ and $j$, and the scalars $J_{ij}$ fully characterize the geometry, range and strength of the interactions.
Hamiltonian evolution is interspersed with a rapid and repeated sequence of $k$ pulses, denoted $P_i$. More specifically, each pulse is followed by free evolution under $H$ for a duration $\tau_i$.
Assuming that the manipulations are sufficiently fast, one can rewrite the unitary evolution (Floquet unitary) over one such $k$-cycle as, 
\begin{align}
\label{eqn:unitary_with_pulse}
    U(T) = e^{-i H\tau_k} P_k \dots e^{-iH\tau_2} P_2 e^{-iH\tau_1}P_1,
\end{align}
where  $T = \sum_{i=1}^k \tau_i$ is the total time duration of the cycle~\footnote{More specifically, we consider a sequence $\{P_i\}$ such that $P_k \dots P_2 P_1 = \mathbb{I}$ by appropriately setting either $P_1$ or $P_k$.}.
At integer multiples of  $T$, the time evolution is captured by an effective Hamiltonian $H_\textrm{eff}$, defined by $U(T) = \exp{\left(-iH_\textrm{eff} T\right)}$.

In the case of both dynamical decoupling and Hamiltonian engineering, the key idea underlying our approach is to design a finite pulse sequence such that $H_\textrm{eff}$ approximates a desired target Hamiltonian.
Defining $U_i \equiv P_i P_{i-1} \dots P_2 P_1$ and $U_0 \equiv \mathbb{I}$, one can rewrite Eq.~\eqref{eqn:unitary_with_pulse} as
\begin{align}
    U(T) = e^{-i \bar{H}_k\tau_k}  \dots e^{-i\bar{H}_2\tau_2} e^{-i\bar{H}_1\tau_1},
\end{align}
where $\bar{H}_i = U^\dagger_i H U_i$.
By moving into this so-called toggling frame \cite{Haeberlen:1968zz}, the pulsed unitary dynamics [Eq.~(2)] can be captured by continuous evolution under a time-dependent Hamiltonian.
For small $T$, a good approximation of $H_\textrm{eff}$ can be obtained using a Magnus expansion~\cite{Mori:2016wb} $H_\textrm{eff} = \sum_{q=0} H_\textrm{eff}^{(q)} $; while our analytics will only consider the leading order  effective Hamiltonian,
\begin{align}
\label{eqn:magnus_zero}
H_\textrm{eff} \approx H_\textrm{eff}^{(0)} = \sum_i \frac{\tau_i}{T} \bar{H}_i,
\end{align}
our numerical computations will simulate the exact time evolution.
So long as $ J_{ij} ||h_{i,j}|| T \ll 1$ for every $i,j$, a low order Magnus expansion can already capture the system's effective dynamics for exponentially long time-scales~\cite{Abanin:2015uh,Abanin:2015uy,Mori:2016wb,Kuwahara:2016dh}.
Also, from the linearity of Eq.~\eqref{eqn:magnus_zero}, we only need to consider a single term $h_{ij}$ and hence omit the qudit indices below.

Consistent with the control available in many-body qudit systems, we focus on the case where one can only apply global single-qudit rotations, i.e., $P_i = p_i^{\otimes N}$ for some $p_i \in SU(d)$.
To represent the interactions, we use a trace orthonormal operator basis  $\{\lambda_\mu \}$ with $\trace{\lambda_\mu \lambda_\nu } = 2\delta_{\mu \nu}$.
In this basis, the most general two-qudit interaction can be written as
\begin{align}
    \label{eqn:c_rep}
    h = \sum_{\mu \nu} C_{\mu \nu} \lambda_\mu \otimes \lambda_\nu.
\end{align}
Hermiticity and the exchange symmetry imply that $C$ is a real symmetric $m \times m$ matrix. For a given $h$, the matrix $C$ can be explicitly  obtained using $C_{\mu \nu} =\trace{ h \lambda_\mu \otimes \lambda_\nu} /4$.
\emph{Interaction Decoupling.}---We now derive a necessary and sufficient condition for the full decoupling (or cancellation) of an interacting qudit Hamiltonian.
\begin{theorem}
\label{theorem:iff_condition}
For a given two-qudit interaction $h$, there exists a finite sequence $\{p_i\} \subset SU(d)$, or equivalently $\{u_i\} \subset SU(d)$, and $\{\tau_i\} \subset \mathbb{R}^+$, such that $h_\textrm{eff} = \sum_i \frac{\tau_i}{T} (u^\dagger_i \otimes u^\dagger_i ) h (u_i \otimes u_i) = 0$ if and only if the $C$ matrix of $h$ is traceless, i.e. $\trace{C} = \sum_\mu \trace{h \lambda_\mu \otimes \lambda_\mu}/4 = 0$. 
\end{theorem}

\begin{proof}
For convenience we work with interactions represented as $C$ matrices, whose 
transformation under a unitary rotation  $u_i\otimes u_i$ is given by,
\begin{align}
\sum_{\mu \nu} C_{\mu \nu} \lambda_\mu \otimes \lambda_\nu 
& \mapsto
\sum_{\mu \nu} 
C_{\mu \nu}  \left( u^\dagger_i \lambda_\mu u_i \right)
    \otimes 
    \left( u^\dagger_i \lambda_\nu u_i\right)\\
&\equiv
\sum_{\mu \nu}
C^{(i)}_{\mu \nu}\lambda_\mu \otimes \lambda_\nu,
\end{align}
where the coefficients $C^{(i)}_{\mu \nu}$ are defined by the equality above.
More specifically, two matrices $C^{(i)}$ and $C$ are related by the transformation $C^{(i)} = \left(O^i\right)^T C O^i$, where $O^{i}_{\nu' \nu} \equiv \frac{1}{2} \trace{ \lambda_\nu u_i^\dagger \lambda_{\nu'} u_i}$.
Taking into account the full sequence of unitary pulses yields the $C$ matrix for the effective Hamiltonian as, 
\begin{align}
\label{eqn:eff_ham_c_rep}
    C_\textrm{eff} = \sum_i \alpha_i \left(O^{i}\right)^T C O^i.
\end{align}
where $\alpha_i = \tau_i / T$ characterizes the relative timescale of the various intermediary free evolutions. 
Intuitively, Eq.~\eqref{eqn:eff_ham_c_rep} demonstrates that the effective interaction is simply given by a weighted average of ``rotated'' versions of the original interaction.
Indeed, it can be easily shown that $O^i$ is a real orthogonal matrix~\cite{supp_info}.

First, one immediately sees that the trace of $C$ is preserved. Thus, from the perspective of interaction decoupling, it is necessary for the original $C$ matrix to be traceless in order for the effective Hamiltonian to be fully decoupled. 
Second, this also naturally suggests a decomposition of  a general interaction into two components: an isotropic part with non-zero trace and a traceless anisotropic piece. 
Since $C$ is a real-symmetric matrix, there exists only one linearly independent isotropic component that is proportional to the identity matrix. The corresponding two-qudit interaction is $h_\textrm{iso} \propto \sum_\mu \lambda_\mu \otimes \lambda_\mu$.
Eq.~\eqref{eqn:eff_ham_c_rep} shows that any isotropic interaction cannot be modified by  global pulses as it is invariant under $SU(d)$ rotations.

To prove the opposite direction (sufficiency), we construct a pulse sequence that explicitly cancels any interaction ($C_\textrm{eff}=0$) given that the interaction is purely anisotropic.
The design principle of this ``universal decoupling'' sequence is simple:  find a finite set of $\{u_i\}$ such that the corresponding $\{O^i\}$ are ``uniformly'' distributed; this  strategy is reminiscent of unitary $2-$designs, but here, we have one additional control knob, corresponding to the choices of $\alpha_i$.
Interestingly, a very related problem has been already studied in quantum information science.
In Ref.~\cite{PhysRevA.61.062313}, D\"ur \emph{et al} introduce a depolarization superoperator $\mathcal{D}$ that acts on a density matrix $\rho$ of a two-qudit system
\begin{align}
    \mathcal{D} (\rho) = 
    A_d \frac{\trace{A_d \rho}}
    {\trace{A_d}}
    +
    S_d \frac{\trace{S_d \rho}}
    {\trace{S_d}},
\end{align}
where $S_d$($A_d$) is the projector onto even(odd) eigenspace
of the exchange operator $\Pi_d = \sum_{i,j=1}^d \ket{ij}\bra{ji}$, i.e., $ A_d = (\mathbb{I}- \Pi_d)/2$ and $S_d = 1- A_d=(\mathbb{I} + \Pi_d)/2$.
It is shown, by explicit construction, that $\mathcal{D}(\cdot)$ can be implemented by a finite sequence of probabilistic bilocal operations, $\sum_{i=1}^{k} p_i \left(v^\dagger_i \otimes v^\dagger_i\right) \rho \left( v_i \otimes v_i \right)= \mathcal{D}(\rho)$,
where $\{p_i\}$ is a probability distribution and $\{ v_i \} \subset SU(d)$.
Here, we re-interpret the super-operator as dynamical decoupling sequence via the mapping: $p_i \rightarrow \alpha_i p_i$ and $v_i \rightarrow u_i $.
To show that this is a universal decoupling sequence, we demonstrate that for an arbitrary interaction $h$, $\trace{S_d h} = -\trace{A_d h} = \trace{C}$; thus, $\trace{C}=0$ implies $\mathcal{D}(h) = 0 $.
The proof is simple: for $h$ acting on qudits $A$ and $B$, 
\begin{align}
    \trace{h \Pi_d} &= \sum_{\mu \nu ij}C_{\mu \nu} \trace{ \lambda_\mu^A \otimes \lambda_\nu^B \ket{ij}\bra{ji} }\\
    &=\sum_{\mu \nu i j}C_{\mu \nu} \bra{j_A}\lambda_\mu^A \ket{i_A} \bra{i_B}\lambda_\nu^B\ket{j_B}\\
    &=\sum_{\mu \nu} C_{\mu \nu} \trace{\lambda_\mu \lambda_\nu} = 2 \trace{C} \label{eqn:proof_3},
\end{align}
where we have explicitly dropped the qudit indices and the tensor product [Eq.~\eqref{eqn:proof_3}] to emphasize that $\lambda_{\mu(\nu)}$ are  matrices.
Finally, noting that $\trace{h} = \sum_{\mu \nu} \trace{\lambda_\mu^A  \otimes \lambda_\nu^B} = 0$, we obtain $\trace{S_d h} = -\trace{A_d h} = \trace{h \Pi_d}/2= \trace{C}$, which completes the proof of Theorem 1. 
\end{proof} 
\emph{Hamiltonian Engineering.}---The previous case of interaction decoupling can be viewed as a specific example of a more general question: given an initial set of interactions $h_0$, a target Hamiltonian $h_f$   and a finite set of available unitaries $\mathcal{U}$, is there a pulse sequence such that, $\sum_i \frac{\tau_i}{T} (u_i^\dagger\otimes u_i^\dagger) h_0 (u_i \otimes u_i)= \beta h_f$ for a constant $\beta>0$? If answered in the affirmative, does there exist an efficient algorithm to construct the desired pulse sequence?   In what follows we describe such an algorithm~\footnote{We consider  $\mathcal{U}$ to be constructed from a set of composite rotations made from simple pulses up to constant depth. This is a particularly natural restriction in the context of experiments, where finite precision limits the available operations. 
Also, noting that the actual pulse to be applied is $p_i = u_i u_{i-1}^\dagger$, we assume that if $u_i$ and $u_{i-1}$ are experimentally feasible, then so is $p_i$.
Finally, a reduction of the interaction strength is inevitable and captured by $\beta$; our algorithm will give the maximum possible value of $\beta$ within the given constraints.}.

\begin{figure}[tb]
\includegraphics[width=3.4in]{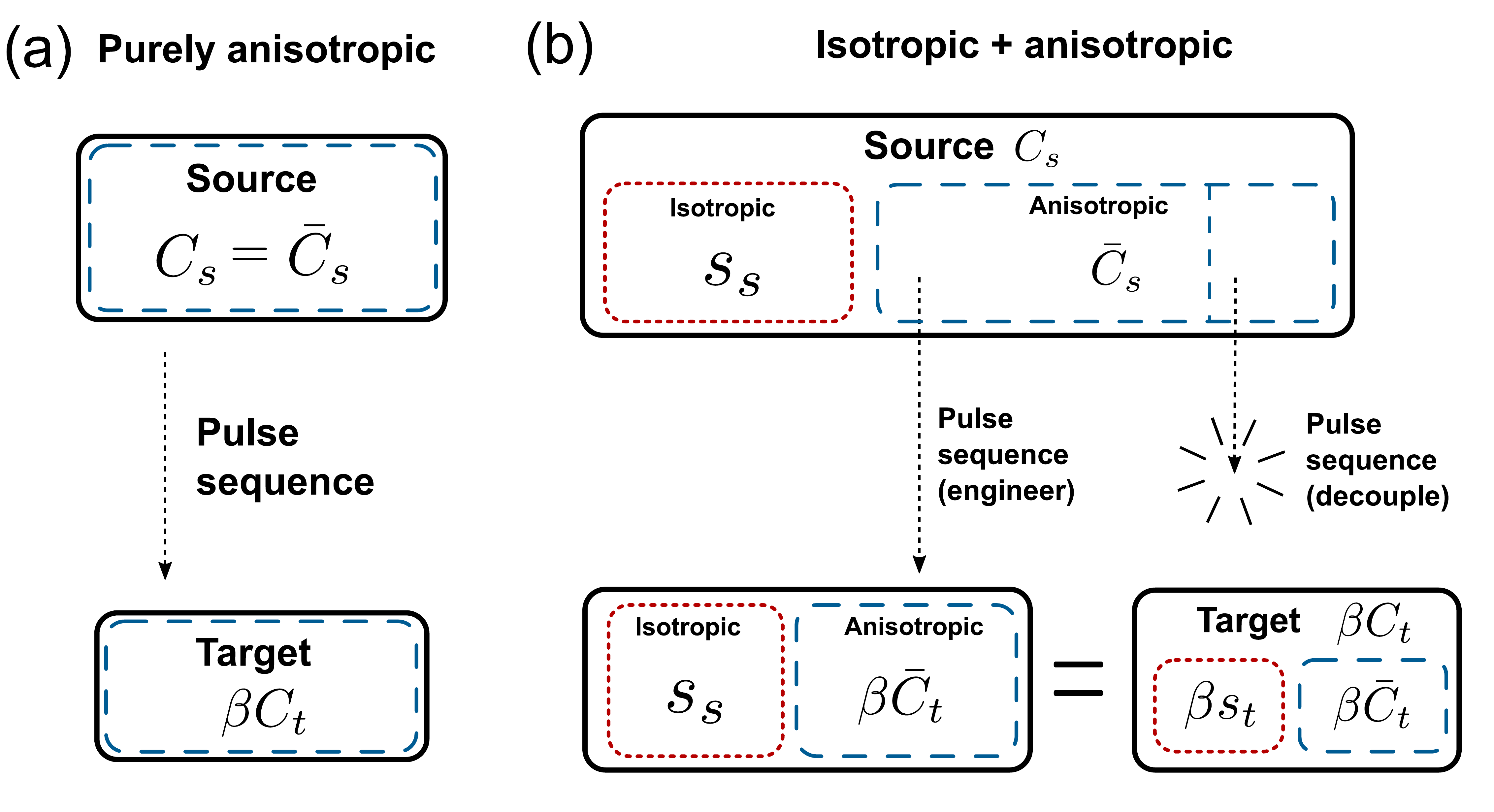}
\caption{ Schematic diagram of interaction engineering. Black solid, red dotted, and blue dashed lines indicate full interactions, isotropic components and anisotropic components, respectively. Dotted arrows represent applications of dynamical decoupling sequence. (a) When both source and target interactions are purely anisotropic ($s_s=s_t=0$), one directly maps interactions. (b) For interactions with both isotropic and anisotropic components, one engineers only the anisotropic component and matches the relative strength by canceling some fraction.} 
\label{fig:decomposition}
\end{figure}

Let us begin by rewriting $h_0$ and $h_f$ in their corresponding $C$ matrices $C_0$ and $C_f$.
We denote the strengths of their isotropic components as $s_0 = \trace{C_0}$ and $s_f = \trace{C_f}$.
As previously discussed, if only one of their $C$ is traceless, $h_0$ cannot be mapped to $h_f$ since the isotropic components can never be decoupled by any pulse sequence.
We will now divide our analysis into two cases: (i) $s_0 = s_f = 0$ and (ii) $s_0, s_f \neq 0$ (Fig.~\ref{fig:decomposition}).

Case (i)  [Fig.~\ref{fig:decomposition}(a)]:  Our strategy is to cancel the portion of the interaction that is \emph{orthogonal} to $C_f$ while maximizing the strength of the remaining piece. 
To illustrate this idea more clearly, we introduce a vector representation of interactions 
\begin{align}
   \left( \vec{w}\right)_a  \equiv \trace{ C \eta_a}/2,
\end{align}
using a matrix basis $\{\eta_a\}$ of dimension $m=d^2-1$.
In this representation, Eq.~\eqref{eqn:eff_ham_c_rep} becomes $\vec{w}_\textrm{eff} = \sum_i \alpha_i M^i \vec{w}$ with $M^i_{ab} \equiv \frac{1}{2} \trace{\eta_a \left(O^i\right)^T \eta_b O^i}$.
Our objective is to maximize $\vec{w}_\textrm{eff} \cdot \vec{w}_f$ while satisfying $\vec{w}_\textrm{eff} \cdot P_\perp = 0$, where $\vec{w}_q$ ($q \in \{0, f\}$) is the vector representation of $C_q$ and $P_\perp$ is the projector on to a space that is orthogonal to $\vec{w}_f$, i.e., $(P_\perp)_{ab}  = \delta_{ab} - (\vec{w}_f)_a (\vec{w}_f)_b/ |\vec{w}_f|^2$.
Interestingly, this task can naturally be cast into the canonical form of Linear Programming, i.e.~maximize $\sum_i \alpha_i  \vec{w}_f \cdot M^i \vec{w}_0$ with respect to $\{\alpha_i\} $ under constraints $\sum \alpha_i P_\perp M^i \vec{w}_0 = 0$, $\sum \alpha_i = 1$, and $\alpha_i \geq 0$ \cite{Bertsimas:1997va}.

Case (ii) [Fig.~\ref{fig:decomposition}(b)]:  In this case, the contributions from the isotropic components cannot be ignored, and they fix the rescaling parameter, $\beta = s_0/s_1$.
Thus, one has to not only engineer the ``shape'' of the anisotropic interaction but also adjust its strength to match with the fixed $\beta$. 
Now our strategy is to decompose the given interaction into three pieces: an isotropic part, a fraction of the anisotropic part to be modified, and the remaining portion to be cancelled.
To this end, one is searching for two pulse sequences, $\mathcal{P}_1=(\{\tau^1_i\}, \{u^1_i\})$, which maps $\bar{C}_0 \mapsto \beta^* \bar{C}_f$ and $\mathcal{P}_2=(\{\tau^2_i\}, \{u^2_i\})$, which cancels $\bar{C}_0 \mapsto 0$. Here,  $\bar{C}_q$ ($q\in \{0, f\}$) is the anisotropic component of $C_q$ and $\beta^*$ is the maximum possible strength. As before, one can use linear programming to efficiently find these sequences. 
If both maps are possible and the engineered interaction strength is sufficiently strong $\beta^* \geq \beta$, one can concatenate two sequences to form $\mathcal{P}_3 = (\{(\beta/\beta^*) \tau^1_i, (1-\beta/\beta^*) \tau^2_i\}, \{ u^1_i, u^2_i\}) $, which maps $C_0\mapsto \beta C_f$.

\begin{figure}[tb]
\includegraphics[width=3.4in]{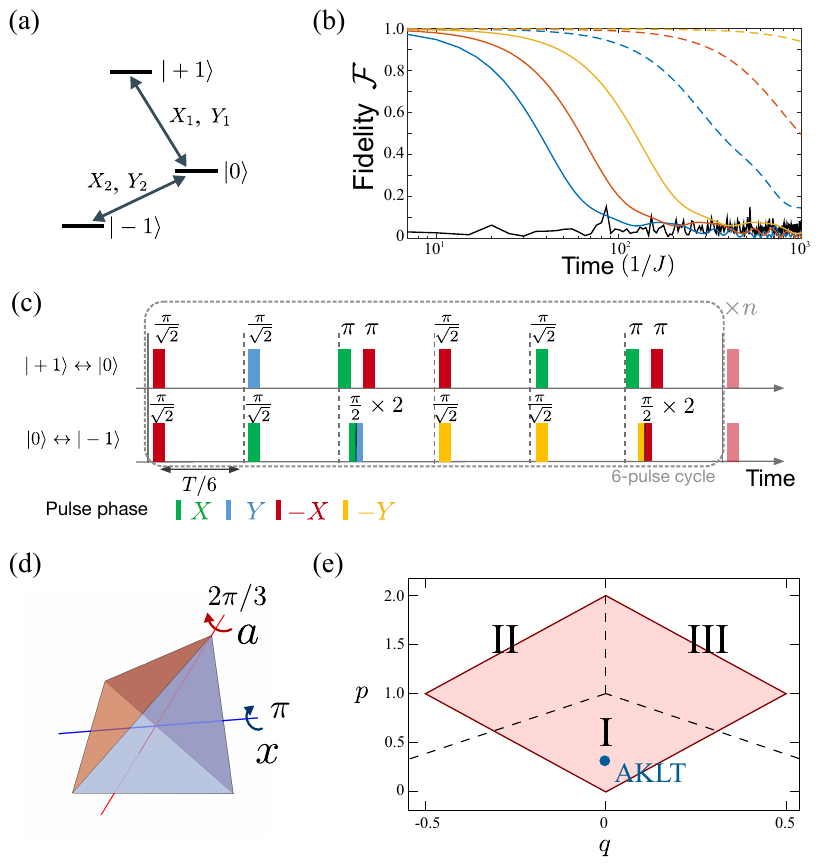}
\caption{{\bf (a)} Level diagram for an anharmonic three level system.  {\bf (b)} Decoupling sequence for spin-1 dipolar interactions. Pulse durations are indicated by rotation angles, and phase choices are color-coded. {\bf (c)} Numerical simulations of decoupling dipolar interactions among $N=6$ spin-1 particles. Black solid line indicates $\mathcal{F}(t)$ in the absence of pulse sequence. Blue, red, and yellow solid lines correspond to $\mathcal{F}(t)$ under a decoupling sequence with $1/JT = 3, 5, 10$, respectively. Dashed lines are for symmetrized sequences.
{\bf(d)} Two generators $\{a, x\}$ of the symmetry group $A_4$.
{\bf(e)} Phase diagram. Three SPT phases (I, II, and III) are distinguished by the transformation of ground state wavefunctions under the action of $a\in A_4$. The colored area indicates the domain of $(p,q)$ that can be engineered from Ising interactions.
Blue dot indicates the AKLT point $(p,q) = (1/3,0)$.  } 
\label{fig:demo}
\end{figure}

\emph{Decoupling spin-1 dipolar interactions.}---We now turn to two examples. 
First, we present a 6-pulse sequence that decouples effective dipole-dipole interactions in an ensemble of spin-1 particles (states $\{\ket{\pm1}, \ket{0}\}$) with  anharmonic level spacings \cite{Kucsko:2016tn},
\begin{align}
    H_d = \sum_{ij} J_{ij}
    &\left[  \sum_{a=1}^{2} (X_{a,i} X_{a,j} + Y_{a,i} Y_{a,j}) \right.\\
    &- \left. (Z_{1,i} + Z_{2,i})(Z_{1,j} + Z_{2,j}) \right], 
\end{align}
where $J_{ij}$ is the interaction strength, while $X_{a,i}$,  $Y_{a,i}$, and $Z_{a,i}$  with $a\in \{1, 2\}$ are generalized Pauli operators for spin transitions between $\ket{0} \leftrightarrow \ket{+1}$ and $\ket{0} \leftrightarrow \ket{-1}$, respectively~[see Fig.~\ref{fig:demo}(a)].
Such a Hamiltonian is ubiquitous in quantum optical systems and arises in the context of ultracold polar molecules, NV centers, and quadrupolar nuclear spins~\cite{Yan:2013fn,Lovchinsky:2017gh,Choi:2016wn}.   
While the solution for the analogous question in dipolar spin-1/2 systems  has been known for a half-century (e.g. WAHUHA), the spin-1 problem  remains an open question.

Motivated by typical experimental constraints, we assume that the available manipulations are limited to a set of composite pulses constructed from up to four $\pm\pi$ or $\pm\frac{\pi}{2}$-pulses between any of the three transitions with two different phases [Fig.~\ref{fig:demo}(a)]. 
Using a simple linear programming algorithm, we find an explicit decoupling sequence using only $6$ pulses $\{P_1, \dots P_6\}$ with equal time durations $\tau_i = T/6$ as depicted in Fig.~\ref{fig:demo}(b).
More detailed explicit expressions for these pulses are provided in Supplementary Material \cite{supp_info}.
In order to test our sequence, we simulate the dynamics of $N=6$ spin-1 particles
with random interaction strengths $J_{ij} \in [-J,J]$ between every pair.
We compute the Floquet unitary $U_T \equiv P_6 e^{-iH_d T/6}P_5 \dots P_1e^{-iH_d T/6}$ and generate stroboscopic time evolution via $(U_T)^n$ with $n\in \mathbb{Z}$. To benchmark the performance of our decoupling sequence, we introduce the fidelity $\mathcal{F}(nT) \equiv |\textrm{tr} \left( (U_T)^n \right)/D|^2$, where $D=3^N$  is the dimension of the Hilbert space.
Since $\mathcal{F}(t) =1$ if and only if the evolution corresponds to the identity unitary, the decay of $\mathcal{F}$  serves as a conservative measure of the performance of our interaction decoupling sequence.%

Figure~\ref{fig:demo}(c) depicts  $\mathcal{F}(t)$ for various values of $T$, demonstrating that the evolution remains trivial up to $\sim 10/J$ for $J T < 1$ (colored solid lines).
Once a given decoupling sequence is found, one can always symmetrize it to further suppress the leading order correction in Magnus expansion~\cite{supp_info}.
In our case, such a sequence involves 10 pulses within the period $2T$. 
Shown as dashed lines in Fig.~\ref{fig:demo} (c), the symmetrized sequence significantly suppresses the interaction for  timescales up to $\sim 100/J$. More generally, it has been rigorously shown that effective dynamics is well captured by low order Magnus expansion up to a long time that scales exponentially in $1/T$~\cite{Abanin:2015uh,Abanin:2015uy,Mori:2016wb,Kuwahara:2016dh}.

\emph{Engineering SPT Hamiltonians.}---As a second example, we show that  a spin-1 chain with nearest neighbor Ising interactions can be directly mapped to a family of SPT Hamiltonians~\cite{supp_info}.
More specifically, given a basic Ising interaction $H_I = \sum_i S_i^z S_{i+1}^z$, one can engineer a two-parameter family of Hamiltonians $H(p,q) = H_1 + p H_2 + q H_3$ with 
\begin{align}
H_1 &= \sum_i \vec{S}_i \cdot \vec{S}_{i+1}, \;\;
H_2 = \sum_i ( \vec{S}_i \cdot \vec{S}_{i+1})^2, \nonumber\\
H_3 &= \sum_i \sum_{(a,b,c)\in S_3}(S_i^a S_{i}^b S_{i+1}^c + S_{i}^a S_{i+1}^b S_{i+1}^c),\nonumber
\end{align}
where $p,q\in \mathbb{R}$,  $\vec{S}_i = (S^x_i, S^y_i, S^z_i)$ is the spin-1 vector operator, and $\sum_{(a,b,c)\in S_3}$ indicates the summation over all permutations of $(x,y,z)$. 
The symmetries of the Hamiltonian include  lattice translation, the bond-centered inversion, and a global internal symmetry $A_4$, which is the symmetry group of a tetrahedron [see Fig~\ref{fig:demo}(d)].
All possible SPT phases protected by these symmetries are explicitly enumerated in Ref.~\cite{Prakash:2016df}.
When $p=1/3$ and $q=0$, the Hamiltonian reduces to celebrated Affleck-Kennedy-Lieb-Tasaki (AKLT) model, whose ground state is exactly solvable and exhibits non-trivial topological edge degrees of freedom \cite{Affleck:1987jy}. 
As $(p,q)$ deviates from this solvable point, phase transitions arise among three distinct regions, I, II, and III, indicated in the numerically obtained phase diagram in Fig.~\ref{fig:demo}(e)~\cite{supp_info}. The ground states in the three phases respect all the symmetries while they are distinguished by the complex $U(1)$ phase that the state picks upon a $120^\circ$ rotation $a\in A_4$ of underlying spins~\cite{supp_info}.
Using our algorithm, we find that $H(p,q)$ with $2|q|\leq p \leq 2 - 2|q|$ can be engineered from $H_I$ [colored area in Fig.~\ref{fig:demo}(e)]. The relative strength of $H(p,q)$ is set to $1/(3+p)$ by isotropic components, and the range of $(p,q)$ is limited by the maximum possible strength of the engineered anisotropic components \cite{supp_info}.
Interestingly, the triple point at $(p,q) = (1,0)$ corresponds to purely isotropic interactions, where the Hamiltonian possesses a larger symmetry group (i.e.~full $SU(d)$).

\emph{Discussions.}---
We now consider the dominant operational imperfections which may arise during the proposed Hamiltonian engineering protocol. First, our periodic driving pulses may cause heating in the many-body system, eventually leading to a featureless infinite temperature state~\cite{DAlessio:2014fg,Lazarides:2014ie,Ponte:2015vj}. Such effects are discussed in Ref.~\cite{Abanin:2015uh,Mori:2016wb,Abanin:2015uy,Kuwahara:2016dh}, and  it has been shown that such energy absorption becomes relevant only after exponentially long times $t^*\sim \exp{[O(1/\bar{J}T)]}$, where $\bar{J} \equiv \max_{i,j} J_{ij} ||h_{ij}||$. A second natural concern is that  our method is based upon engineering the leading order Magnus Hamiltonian $H_\textrm{eff}^{(0)}$, which provides only an approximate description of the full many-body dynamics. 
However, for gapped Hamiltonians, one expects that higher order terms in the Magnus expansion are strongly suppressed so long as $\bar{J}T \ll 1$, implying that the phase should remain stable.
Finally, adiabatic change of parameters can be used to prepare the system in a low-entropy state close to the ground state of the effective Hamiltonian.

Interestingly, the decoupling of interactions may result in dynamical quantum phase transitions for isolated, weakly disordered systems~\cite{DIMBLChoi:2017wm}. In such cases, the interplay of weak disorder, suppressed interactions, and an exponentially slow heating rate can lead to many-body localization, where initial state memories survive for extremely long times. Harnessing these effects may enable the coherent manipulation and storage of quantum information in an interacting many-body system~\cite{QCMBLChoi:2015,MBLStateTransferYao:2015}.

\begin{acknowledgments}
The authors would like to thank H. Zhou, J. Choi, V. Khemani, A. Prakash, J. Haah, A. Gorshkov, Y. Moon, and  J. Taylor for useful discussions. This work was supported through NSF, CUA, the Vannevar Bush Faculty Fellowship, AFOSR Muri and Moore Foundation. S.~C. is supported by Kwanjeong Educational Foundation.
\end{acknowledgments}

\bibliography{refs}

\begin{thebibliography}{51}%
\makeatletter
\providecommand \@ifxundefined [1]{%
 \@ifx{#1\undefined}
}%
\providecommand \@ifnum [1]{%
 \ifnum #1\expandafter \@firstoftwo
 \else \expandafter \@secondoftwo
 \fi
}%
\providecommand \@ifx [1]{%
 \ifx #1\expandafter \@firstoftwo
 \else \expandafter \@secondoftwo
 \fi
}%
\providecommand \natexlab [1]{#1}%
\providecommand \enquote  [1]{``#1''}%
\providecommand \bibnamefont  [1]{#1}%
\providecommand \bibfnamefont [1]{#1}%
\providecommand \citenamefont [1]{#1}%
\providecommand \href@noop [0]{\@secondoftwo}%
\providecommand \href [0]{\begingroup \@sanitize@url \@href}%
\providecommand \@href[1]{\@@startlink{#1}\@@href}%
\providecommand \@@href[1]{\endgroup#1\@@endlink}%
\providecommand \@sanitize@url [0]{\catcode `\\12\catcode `\$12\catcode
  `\&12\catcode `\#12\catcode `\^12\catcode `\_12\catcode `\%12\relax}%
\providecommand \@@startlink[1]{}%
\providecommand \@@endlink[0]{}%
\providecommand \url  [0]{\begingroup\@sanitize@url \@url }%
\providecommand \@url [1]{\endgroup\@href {#1}{\urlprefix }}%
\providecommand \urlprefix  [0]{URL }%
\providecommand \Eprint [0]{\href }%
\providecommand \doibase [0]{http://dx.doi.org/}%
\providecommand \selectlanguage [0]{\@gobble}%
\providecommand \bibinfo  [0]{\@secondoftwo}%
\providecommand \bibfield  [0]{\@secondoftwo}%
\providecommand \translation [1]{[#1]}%
\providecommand \BibitemOpen [0]{}%
\providecommand \bibitemStop [0]{}%
\providecommand \bibitemNoStop [0]{.\EOS\space}%
\providecommand \EOS [0]{\spacefactor3000\relax}%
\providecommand \BibitemShut  [1]{\csname bibitem#1\endcsname}%
\let\auto@bib@innerbib\@empty
%</preamble>
\bibitem [{\citenamefont {Hahn}(1950)}]{Hahn_echo:1950ge}%
  \BibitemOpen
  \bibfield  {author} {\bibinfo {author} {\bibfnamefont {E.~L.}\ \bibnamefont
  {Hahn}},\ }\href@noop {} {\bibfield  {journal} {\bibinfo  {journal} {Physical
  Review}\ }\textbf {\bibinfo {volume} {80}},\ \bibinfo {pages} {580} (\bibinfo
  {year} {1950})}\BibitemShut {NoStop}%
\bibitem [{\citenamefont {Uhrig}(2007)}]{Uhrig:2007ek}%
  \BibitemOpen
  \bibfield  {author} {\bibinfo {author} {\bibfnamefont {G.~S.}\ \bibnamefont
  {Uhrig}},\ }\href@noop {} {\bibfield  {journal} {\bibinfo  {journal}
  {Physical Review Letters}\ }\textbf {\bibinfo {volume} {98}},\ \bibinfo
  {pages} {100504} (\bibinfo {year} {2007})}\BibitemShut {NoStop}%
\bibitem [{\citenamefont {de~Lange}\ \emph {et~al.}(2010)\citenamefont
  {de~Lange}, \citenamefont {Wang}, \citenamefont {Rist{\`e}}, \citenamefont
  {Dobrovitski},\ and\ \citenamefont {Hanson}}]{deLange:2010ga}%
  \BibitemOpen
  \bibfield  {author} {\bibinfo {author} {\bibfnamefont {G.}~\bibnamefont
  {de~Lange}}, \bibinfo {author} {\bibfnamefont {Z.~H.}\ \bibnamefont {Wang}},
  \bibinfo {author} {\bibfnamefont {D.}~\bibnamefont {Rist{\`e}}}, \bibinfo
  {author} {\bibfnamefont {V.~V.}\ \bibnamefont {Dobrovitski}}, \ and\ \bibinfo
  {author} {\bibfnamefont {R.}~\bibnamefont {Hanson}},\ }\href@noop {}
  {\bibfield  {journal} {\bibinfo  {journal} {Science}\ }\textbf {\bibinfo
  {volume} {330}},\ \bibinfo {pages} {60} (\bibinfo {year} {2010})}\BibitemShut
  {NoStop}%
\bibitem [{\citenamefont {Kuo}\ and\ \citenamefont
  {Lidar}(2011)}]{QDDproofLidar:2011aa}%
  \BibitemOpen
  \bibfield  {author} {\bibinfo {author} {\bibfnamefont {W.-J.}\ \bibnamefont
  {Kuo}}\ and\ \bibinfo {author} {\bibfnamefont {D.~A.}\ \bibnamefont
  {Lidar}},\ }\href@noop {} {\bibfield  {journal} {\bibinfo  {journal}
  {Physical Review A}\ }\textbf {\bibinfo {volume} {84}},\ \bibinfo {pages}
  {042329} (\bibinfo {year} {2011})}\BibitemShut {NoStop}%
\bibitem [{\citenamefont {Jiang}\ and\ \citenamefont
  {Imambekov}(2011)}]{UDDproofJiang:2011aa}%
  \BibitemOpen
  \bibfield  {author} {\bibinfo {author} {\bibfnamefont {L.}~\bibnamefont
  {Jiang}}\ and\ \bibinfo {author} {\bibfnamefont {A.}~\bibnamefont
  {Imambekov}},\ }\href@noop {} {\bibfield  {journal} {\bibinfo  {journal}
  {Physical Review A}\ }\textbf {\bibinfo {volume} {84}},\ \bibinfo {pages}
  {060302} (\bibinfo {year} {2011})}\BibitemShut {NoStop}%
\bibitem [{\citenamefont {Maurer}\ \emph {et~al.}(2012)\citenamefont {Maurer},
  \citenamefont {Kucsko}, \citenamefont {Latta}, \citenamefont {Jiang},\ and\
  \citenamefont {Yao}}]{NV_memory:2012Maurer}%
  \BibitemOpen
  \bibfield  {author} {\bibinfo {author} {\bibfnamefont {P.~C.}\ \bibnamefont
  {Maurer}}, \bibinfo {author} {\bibfnamefont {G.}~\bibnamefont {Kucsko}},
  \bibinfo {author} {\bibfnamefont {C.}~\bibnamefont {Latta}}, \bibinfo
  {author} {\bibfnamefont {L.}~\bibnamefont {Jiang}}, \ and\ \bibinfo {author}
  {\bibfnamefont {N.~Y.}\ \bibnamefont {Yao}},\ }\href@noop {} {\bibfield
  {journal} {\bibinfo  {journal} {Science}\ }\textbf {\bibinfo {volume}
  {336}},\ \bibinfo {pages} {1283} (\bibinfo {year} {2012})}\BibitemShut
  {NoStop}%
\bibitem [{\citenamefont {Paz-Silva}\ and\ \citenamefont
  {Lidar}(2013)}]{PazSilva:2013hx}%
  \BibitemOpen
  \bibfield  {author} {\bibinfo {author} {\bibfnamefont {G.~A.}\ \bibnamefont
  {Paz-Silva}}\ and\ \bibinfo {author} {\bibfnamefont {D.~A.}\ \bibnamefont
  {Lidar}},\ }\href@noop {} {\bibfield  {journal} {\bibinfo  {journal}
  {Scientific reports}\ } (\bibinfo {year} {2013})}\BibitemShut {NoStop}%
\bibitem [{\citenamefont {Kessler}\ \emph {et~al.}(2014)\citenamefont
  {Kessler}, \citenamefont {Komar}, \citenamefont {Bishof}, \citenamefont
  {Jiang},\ and\ \citenamefont {S{\o}rensen}}]{entangledQubitClock2014}%
  \BibitemOpen
  \bibfield  {author} {\bibinfo {author} {\bibfnamefont {E.~M.}\ \bibnamefont
  {Kessler}}, \bibinfo {author} {\bibfnamefont {P.}~\bibnamefont {Komar}},
  \bibinfo {author} {\bibfnamefont {M.}~\bibnamefont {Bishof}}, \bibinfo
  {author} {\bibfnamefont {L.}~\bibnamefont {Jiang}}, \ and\ \bibinfo {author}
  {\bibfnamefont {A.~S.}\ \bibnamefont {S{\o}rensen}},\ }\href@noop {}
  {\bibfield  {journal} {\bibinfo  {journal} {Physical Review}\ }\textbf
  {\bibinfo {volume} {112}},\ \bibinfo {pages} {190403} (\bibinfo {year}
  {2014})}\BibitemShut {NoStop}%
\bibitem [{\citenamefont {Lovchinsky}\ \emph {et~al.}(2016)\citenamefont
  {Lovchinsky}, \citenamefont {Sushkov}, \citenamefont {Urbach}, \citenamefont
  {de~Leon}, \citenamefont {Choi}, \citenamefont {De~Greve}, \citenamefont
  {Evans}, \citenamefont {Gertner}, \citenamefont {Bersin}, \citenamefont
  {M{\"u}ller}, \citenamefont {McGuinness}, \citenamefont {Jelezko},
  \citenamefont {Walsworth}, \citenamefont {Park},\ and\ \citenamefont
  {Lukin}}]{Lovchinsky:2016dq}%
  \BibitemOpen
  \bibfield  {author} {\bibinfo {author} {\bibfnamefont {I.}~\bibnamefont
  {Lovchinsky}}, \bibinfo {author} {\bibfnamefont {A.~O.}\ \bibnamefont
  {Sushkov}}, \bibinfo {author} {\bibfnamefont {E.}~\bibnamefont {Urbach}},
  \bibinfo {author} {\bibfnamefont {N.~P.}\ \bibnamefont {de~Leon}}, \bibinfo
  {author} {\bibfnamefont {S.}~\bibnamefont {Choi}}, \bibinfo {author}
  {\bibfnamefont {K.}~\bibnamefont {De~Greve}}, \bibinfo {author}
  {\bibfnamefont {R.}~\bibnamefont {Evans}}, \bibinfo {author} {\bibfnamefont
  {R.}~\bibnamefont {Gertner}}, \bibinfo {author} {\bibfnamefont
  {E.}~\bibnamefont {Bersin}}, \bibinfo {author} {\bibfnamefont
  {C.}~\bibnamefont {M{\"u}ller}}, \bibinfo {author} {\bibfnamefont
  {L.}~\bibnamefont {McGuinness}}, \bibinfo {author} {\bibfnamefont
  {F.}~\bibnamefont {Jelezko}}, \bibinfo {author} {\bibfnamefont {R.~L.}\
  \bibnamefont {Walsworth}}, \bibinfo {author} {\bibfnamefont {H.}~\bibnamefont
  {Park}}, \ and\ \bibinfo {author} {\bibfnamefont {M.~D.}\ \bibnamefont
  {Lukin}},\ }\href@noop {} {\bibfield  {journal} {\bibinfo  {journal}
  {Science}\ }\textbf {\bibinfo {volume} {351}},\ \bibinfo {pages} {836}
  (\bibinfo {year} {2016})}\BibitemShut {NoStop}%
\bibitem [{\citenamefont {Lovchinsky}\ \emph {et~al.}(2017)\citenamefont
  {Lovchinsky}, \citenamefont {Sanchez-Yamagishi}, \citenamefont {Urbach},
  \citenamefont {Choi}, \citenamefont {Fang}, \citenamefont {Andersen},
  \citenamefont {Watanabe}, \citenamefont {Taniguchi}, \citenamefont
  {Bylinskii}, \citenamefont {Kaxiras}, \citenamefont {Kim}, \citenamefont
  {Park},\ and\ \citenamefont {Lukin}}]{Lovchinsky:2017gh}%
  \BibitemOpen
  \bibfield  {author} {\bibinfo {author} {\bibfnamefont {I.}~\bibnamefont
  {Lovchinsky}}, \bibinfo {author} {\bibfnamefont {J.~D.}\ \bibnamefont
  {Sanchez-Yamagishi}}, \bibinfo {author} {\bibfnamefont {E.~K.}\ \bibnamefont
  {Urbach}}, \bibinfo {author} {\bibfnamefont {S.}~\bibnamefont {Choi}},
  \bibinfo {author} {\bibfnamefont {S.}~\bibnamefont {Fang}}, \bibinfo {author}
  {\bibfnamefont {T.~I.}\ \bibnamefont {Andersen}}, \bibinfo {author}
  {\bibfnamefont {K.}~\bibnamefont {Watanabe}}, \bibinfo {author}
  {\bibfnamefont {T.}~\bibnamefont {Taniguchi}}, \bibinfo {author}
  {\bibfnamefont {A.}~\bibnamefont {Bylinskii}}, \bibinfo {author}
  {\bibfnamefont {E.}~\bibnamefont {Kaxiras}}, \bibinfo {author} {\bibfnamefont
  {P.}~\bibnamefont {Kim}}, \bibinfo {author} {\bibfnamefont {H.}~\bibnamefont
  {Park}}, \ and\ \bibinfo {author} {\bibfnamefont {M.~D.}\ \bibnamefont
  {Lukin}},\ }\href@noop {} {\bibfield  {journal} {\bibinfo  {journal}
  {Science}\ }\textbf {\bibinfo {volume} {355}},\ \bibinfo {pages} {503}
  (\bibinfo {year} {2017})}\BibitemShut {NoStop}%
\bibitem [{\citenamefont {Lindner}\ \emph {et~al.}(2011)\citenamefont
  {Lindner}, \citenamefont {Refael},\ and\ \citenamefont
  {Galitski}}]{Lindner:2011ip}%
  \BibitemOpen
  \bibfield  {author} {\bibinfo {author} {\bibfnamefont {N.~H.}\ \bibnamefont
  {Lindner}}, \bibinfo {author} {\bibfnamefont {G.}~\bibnamefont {Refael}}, \
  and\ \bibinfo {author} {\bibfnamefont {V.}~\bibnamefont {Galitski}},\
  }\href@noop {} {\bibfield  {journal} {\bibinfo  {journal} {Nature Physics}\
  }\textbf {\bibinfo {volume} {7}},\ \bibinfo {pages} {490} (\bibinfo {year}
  {2011})}\BibitemShut {NoStop}%
\bibitem [{\citenamefont {Jiang}\ \emph {et~al.}(2011)\citenamefont {Jiang},
  \citenamefont {Kitagawa}, \citenamefont {Alicea}, \citenamefont {Akhmerov},
  \citenamefont {Pekker}, \citenamefont {Refael}, \citenamefont {Cirac},
  \citenamefont {Demler}, \citenamefont {Lukin},\ and\ \citenamefont
  {Zoller}}]{FloquetMajoranaJiang:2011aa}%
  \BibitemOpen
  \bibfield  {author} {\bibinfo {author} {\bibfnamefont {L.}~\bibnamefont
  {Jiang}}, \bibinfo {author} {\bibfnamefont {T.}~\bibnamefont {Kitagawa}},
  \bibinfo {author} {\bibfnamefont {J.}~\bibnamefont {Alicea}}, \bibinfo
  {author} {\bibfnamefont {A.~R.}\ \bibnamefont {Akhmerov}}, \bibinfo {author}
  {\bibfnamefont {D.}~\bibnamefont {Pekker}}, \bibinfo {author} {\bibfnamefont
  {G.}~\bibnamefont {Refael}}, \bibinfo {author} {\bibfnamefont {J.~I.}\
  \bibnamefont {Cirac}}, \bibinfo {author} {\bibfnamefont {E.}~\bibnamefont
  {Demler}}, \bibinfo {author} {\bibfnamefont {M.~D.}\ \bibnamefont {Lukin}}, \
  and\ \bibinfo {author} {\bibfnamefont {P.}~\bibnamefont {Zoller}},\
  }\href@noop {} {\bibfield  {journal} {\bibinfo  {journal} {Physical Review
  Letters}\ }\textbf {\bibinfo {volume} {106}},\ \bibinfo {pages} {220402}
  (\bibinfo {year} {2011})}\BibitemShut {NoStop}%
\bibitem [{\citenamefont {Iadecola}\ \emph {et~al.}(2015)\citenamefont
  {Iadecola}, \citenamefont {Santos},\ and\ \citenamefont
  {Chamon}}]{Iadecola:2015dg}%
  \BibitemOpen
  \bibfield  {author} {\bibinfo {author} {\bibfnamefont {T.}~\bibnamefont
  {Iadecola}}, \bibinfo {author} {\bibfnamefont {L.~H.}\ \bibnamefont
  {Santos}}, \ and\ \bibinfo {author} {\bibfnamefont {C.}~\bibnamefont
  {Chamon}},\ }\href@noop {} {\bibfield  {journal} {\bibinfo  {journal}
  {Physical Review B}\ }\textbf {\bibinfo {volume} {92}},\ \bibinfo {pages}
  {125107} (\bibinfo {year} {2015})}\BibitemShut {NoStop}%
\bibitem [{\citenamefont {Khemani}\ \emph {et~al.}(2016)\citenamefont
  {Khemani}, \citenamefont {Lazarides}, \citenamefont {Moessner},\ and\
  \citenamefont {Sondhi}}]{Khemani:2015gd}%
  \BibitemOpen
  \bibfield  {author} {\bibinfo {author} {\bibfnamefont {V.}~\bibnamefont
  {Khemani}}, \bibinfo {author} {\bibfnamefont {A.}~\bibnamefont {Lazarides}},
  \bibinfo {author} {\bibfnamefont {R.}~\bibnamefont {Moessner}}, \ and\
  \bibinfo {author} {\bibfnamefont {S.~L.}\ \bibnamefont {Sondhi}},\
  }\href@noop {} {\bibfield  {journal} {\bibinfo  {journal} {Physical Review
  Letters}\ }\textbf {\bibinfo {volume} {116}},\ \bibinfo {pages} {250401}
  (\bibinfo {year} {2016})}\BibitemShut {NoStop}%
\bibitem [{\citenamefont {Else}\ \emph {et~al.}(2016)\citenamefont {Else},
  \citenamefont {Bauer},\ and\ \citenamefont {Nayak}}]{Else:2016gf}%
  \BibitemOpen
  \bibfield  {author} {\bibinfo {author} {\bibfnamefont {D.~V.}\ \bibnamefont
  {Else}}, \bibinfo {author} {\bibfnamefont {B.}~\bibnamefont {Bauer}}, \ and\
  \bibinfo {author} {\bibfnamefont {C.}~\bibnamefont {Nayak}},\ }\href@noop {}
  {\bibfield  {journal} {\bibinfo  {journal} {Physical Review Letters}\
  }\textbf {\bibinfo {volume} {117}},\ \bibinfo {pages} {090402} (\bibinfo
  {year} {2016})}\BibitemShut {NoStop}%
\bibitem [{\citenamefont {von Keyserlingk}\ \emph {et~al.}(2016)\citenamefont
  {von Keyserlingk}, \citenamefont {Khemani},\ and\ \citenamefont
  {Sondhi}}]{vonKeyserlingk:2016ev}%
  \BibitemOpen
  \bibfield  {author} {\bibinfo {author} {\bibfnamefont {C.~W.}\ \bibnamefont
  {von Keyserlingk}}, \bibinfo {author} {\bibfnamefont {V.}~\bibnamefont
  {Khemani}}, \ and\ \bibinfo {author} {\bibfnamefont {S.~L.}\ \bibnamefont
  {Sondhi}},\ }\href@noop {} {\bibfield  {journal} {\bibinfo  {journal}
  {Physical Review B}\ }\textbf {\bibinfo {volume} {94}},\ \bibinfo {pages}
  {085112} (\bibinfo {year} {2016})}\BibitemShut {NoStop}%
\bibitem [{\citenamefont {Yao}\ \emph {et~al.}(2017)\citenamefont {Yao},
  \citenamefont {Potter}, \citenamefont {Potirniche},\ and\ \citenamefont
  {Vishwanath}}]{Yao_dtc:2016wp}%
  \BibitemOpen
  \bibfield  {author} {\bibinfo {author} {\bibfnamefont {N.~Y.}\ \bibnamefont
  {Yao}}, \bibinfo {author} {\bibfnamefont {A.~C.}\ \bibnamefont {Potter}},
  \bibinfo {author} {\bibfnamefont {I.-D.}\ \bibnamefont {Potirniche}}, \ and\
  \bibinfo {author} {\bibfnamefont {A.}~\bibnamefont {Vishwanath}},\
  }\href@noop {} {\bibfield  {journal} {\bibinfo  {journal} {Physical Review
  Letters}\ }\textbf {\bibinfo {volume} {118}},\ \bibinfo {pages} {030401}
  (\bibinfo {year} {2017})}\BibitemShut {NoStop}%
\bibitem [{\citenamefont {Haeberlen}\ and\ \citenamefont
  {Waugh}(1968)}]{Haeberlen:1968zz}%
  \BibitemOpen
  \bibfield  {author} {\bibinfo {author} {\bibfnamefont {U.}~\bibnamefont
  {Haeberlen}}\ and\ \bibinfo {author} {\bibfnamefont {J.~S.}\ \bibnamefont
  {Waugh}},\ }\href@noop {} {\bibfield  {journal} {\bibinfo  {journal} {Phys.
  Rev.}\ }\textbf {\bibinfo {volume} {175}},\ \bibinfo {pages} {453} (\bibinfo
  {year} {1968})}\BibitemShut {NoStop}%
\bibitem [{\citenamefont {Waugh}\ \emph {et~al.}(1968)\citenamefont {Waugh},
  \citenamefont {Huber},\ and\ \citenamefont {Haeberlen}}]{Waugh:1968im}%
  \BibitemOpen
  \bibfield  {author} {\bibinfo {author} {\bibfnamefont {J.~S.}\ \bibnamefont
  {Waugh}}, \bibinfo {author} {\bibfnamefont {L.~M.}\ \bibnamefont {Huber}}, \
  and\ \bibinfo {author} {\bibfnamefont {U.}~\bibnamefont {Haeberlen}},\
  }\href@noop {} {\bibfield  {journal} {\bibinfo  {journal} {Physical Review
  Letters}\ }\textbf {\bibinfo {volume} {20}},\ \bibinfo {pages} {180}
  (\bibinfo {year} {1968})}\BibitemShut {NoStop}%
\bibitem [{\citenamefont {Yan}\ \emph {et~al.}(2013)\citenamefont {Yan},
  \citenamefont {Moses}, \citenamefont {Gadway}, \citenamefont {Covey},
  \citenamefont {Hazzard}, \citenamefont {Rey}, \citenamefont {Jin},\ and\
  \citenamefont {Ye}}]{Yan:2013fn}%
  \BibitemOpen
  \bibfield  {author} {\bibinfo {author} {\bibfnamefont {B.}~\bibnamefont
  {Yan}}, \bibinfo {author} {\bibfnamefont {S.~A.}\ \bibnamefont {Moses}},
  \bibinfo {author} {\bibfnamefont {B.}~\bibnamefont {Gadway}}, \bibinfo
  {author} {\bibfnamefont {J.~P.}\ \bibnamefont {Covey}}, \bibinfo {author}
  {\bibfnamefont {K.~R.~A.}\ \bibnamefont {Hazzard}}, \bibinfo {author}
  {\bibfnamefont {A.~M.}\ \bibnamefont {Rey}}, \bibinfo {author} {\bibfnamefont
  {D.~S.}\ \bibnamefont {Jin}}, \ and\ \bibinfo {author} {\bibfnamefont
  {J.}~\bibnamefont {Ye}},\ }\href@noop {} {\bibfield  {journal} {\bibinfo
  {journal} {Nature}\ }\textbf {\bibinfo {volume} {501}},\ \bibinfo {pages}
  {521} (\bibinfo {year} {2013})}\BibitemShut {NoStop}%
\bibitem [{\citenamefont {Brinkmann}\ and\ \citenamefont
  {Ed{\'e}n}(2004)}]{Brinkmann:2004fk}%
  \BibitemOpen
  \bibfield  {author} {\bibinfo {author} {\bibfnamefont {A.}~\bibnamefont
  {Brinkmann}}\ and\ \bibinfo {author} {\bibfnamefont {M.}~\bibnamefont
  {Ed{\'e}n}},\ }\href@noop {} {\bibfield  {journal} {\bibinfo  {journal} {The
  Journal of chemical physics}\ }\textbf {\bibinfo {volume} {120}},\ \bibinfo
  {pages} {11726} (\bibinfo {year} {2004})}\BibitemShut {NoStop}%
\bibitem [{\citenamefont {Ajoy}\ and\ \citenamefont
  {Cappellaro}(2013)}]{Ajoy:2013hf}%
  \BibitemOpen
  \bibfield  {author} {\bibinfo {author} {\bibfnamefont {A.}~\bibnamefont
  {Ajoy}}\ and\ \bibinfo {author} {\bibfnamefont {P.}~\bibnamefont
  {Cappellaro}},\ }\href@noop {} {\bibfield  {journal} {\bibinfo  {journal}
  {Physical Review Letters}\ }\textbf {\bibinfo {volume} {110}},\ \bibinfo
  {pages} {220503} (\bibinfo {year} {2013})}\BibitemShut {NoStop}%
\bibitem [{\citenamefont {Frydrych}\ \emph {et~al.}(2014)\citenamefont
  {Frydrych}, \citenamefont {Alber},\ and\ \citenamefont {Ba{\v
  z}ant}}]{Frydrych:2014il}%
  \BibitemOpen
  \bibfield  {author} {\bibinfo {author} {\bibfnamefont {H.}~\bibnamefont
  {Frydrych}}, \bibinfo {author} {\bibfnamefont {G.}~\bibnamefont {Alber}}, \
  and\ \bibinfo {author} {\bibfnamefont {P.}~\bibnamefont {Ba{\v z}ant}},\
  }\href@noop {} {\bibfield  {journal} {\bibinfo  {journal} {Physical Review
  A}\ }\textbf {\bibinfo {volume} {89}},\ \bibinfo {pages} {022320} (\bibinfo
  {year} {2014})}\BibitemShut {NoStop}%
\bibitem [{\citenamefont {Hayes}\ \emph {et~al.}(2014)\citenamefont {Hayes},
  \citenamefont {Flammia},\ and\ \citenamefont {Biercuk}}]{Hayes:2014vj}%
  \BibitemOpen
  \bibfield  {author} {\bibinfo {author} {\bibfnamefont {D.}~\bibnamefont
  {Hayes}}, \bibinfo {author} {\bibfnamefont {S.~T.}\ \bibnamefont {Flammia}},
  \ and\ \bibinfo {author} {\bibfnamefont {M.~J.}\ \bibnamefont {Biercuk}},\
  }\href@noop {} {\bibfield  {journal} {\bibinfo  {journal} {New Journal of
  Physics}\ } (\bibinfo {year} {2014})}\BibitemShut {NoStop}%
\bibitem [{\citenamefont {Senko}\ \emph {et~al.}(2015)\citenamefont {Senko},
  \citenamefont {Richerme}, \citenamefont {Smith}, \citenamefont {Lee},
  \citenamefont {Cohen}, \citenamefont {Retzker},\ and\ \citenamefont
  {Monroe}}]{Senko:2015hf}%
  \BibitemOpen
  \bibfield  {author} {\bibinfo {author} {\bibfnamefont {C.}~\bibnamefont
  {Senko}}, \bibinfo {author} {\bibfnamefont {P.}~\bibnamefont {Richerme}},
  \bibinfo {author} {\bibfnamefont {J.}~\bibnamefont {Smith}}, \bibinfo
  {author} {\bibfnamefont {A.}~\bibnamefont {Lee}}, \bibinfo {author}
  {\bibfnamefont {I.}~\bibnamefont {Cohen}}, \bibinfo {author} {\bibfnamefont
  {A.}~\bibnamefont {Retzker}}, \ and\ \bibinfo {author} {\bibfnamefont
  {C.}~\bibnamefont {Monroe}},\ }\href@noop {} {\bibfield  {journal} {\bibinfo
  {journal} {Physical Review X}\ }\textbf {\bibinfo {volume} {5}},\ \bibinfo
  {pages} {021026} (\bibinfo {year} {2015})}\BibitemShut {NoStop}%
\bibitem [{\citenamefont {Choi}\ \emph
  {et~al.}(2017{\natexlab{a}})\citenamefont {Choi}, \citenamefont {Choi},
  \citenamefont {Landig}, \citenamefont {Kucsko}, \citenamefont {Zhou},
  \citenamefont {Isoya}, \citenamefont {Jelezko}, \citenamefont {Onoda},
  \citenamefont {Sumiya}, \citenamefont {Khemani}, \citenamefont {von
  Keyserlingk}, \citenamefont {Yao}, \citenamefont {Demler},\ and\
  \citenamefont {Lukin}}]{Choi:2016wn}%
  \BibitemOpen
  \bibfield  {author} {\bibinfo {author} {\bibfnamefont {S.}~\bibnamefont
  {Choi}}, \bibinfo {author} {\bibfnamefont {J.}~\bibnamefont {Choi}}, \bibinfo
  {author} {\bibfnamefont {R.}~\bibnamefont {Landig}}, \bibinfo {author}
  {\bibfnamefont {G.}~\bibnamefont {Kucsko}}, \bibinfo {author} {\bibfnamefont
  {H.}~\bibnamefont {Zhou}}, \bibinfo {author} {\bibfnamefont {J.}~\bibnamefont
  {Isoya}}, \bibinfo {author} {\bibfnamefont {F.}~\bibnamefont {Jelezko}},
  \bibinfo {author} {\bibfnamefont {S.}~\bibnamefont {Onoda}}, \bibinfo
  {author} {\bibfnamefont {H.}~\bibnamefont {Sumiya}}, \bibinfo {author}
  {\bibfnamefont {V.}~\bibnamefont {Khemani}}, \bibinfo {author} {\bibfnamefont
  {C.}~\bibnamefont {von Keyserlingk}}, \bibinfo {author} {\bibfnamefont
  {N.~Y.}\ \bibnamefont {Yao}}, \bibinfo {author} {\bibfnamefont {E.~A.}\
  \bibnamefont {Demler}}, \ and\ \bibinfo {author} {\bibfnamefont {M.~D.}\
  \bibnamefont {Lukin}},\ }\href@noop {} {\bibfield  {journal} {\bibinfo
  {journal} {Nature}\ }\textbf {\bibinfo {volume} {543}},\ \bibinfo {pages}
  {221} (\bibinfo {year} {2017}{\natexlab{a}})}\BibitemShut {NoStop}%
\bibitem [{\citenamefont {Potts}\ and\ \citenamefont
  {Domb}(2008)}]{Potts:1951be}%
  \BibitemOpen
  \bibfield  {author} {\bibinfo {author} {\bibfnamefont {R.~B.}\ \bibnamefont
  {Potts}}\ and\ \bibinfo {author} {\bibfnamefont {C.}~\bibnamefont {Domb}},\
  }\href@noop {} {\bibfield  {journal} {\bibinfo  {journal} {Mathematical
  Proceedings of the Cambridge Philosophical Society}\ }\textbf {\bibinfo
  {volume} {48}},\ \bibinfo {pages} {106} (\bibinfo {year} {2008})}\BibitemShut
  {NoStop}%
\bibitem [{\citenamefont {Huse}(1981)}]{Huse:1981gg}%
  \BibitemOpen
  \bibfield  {author} {\bibinfo {author} {\bibfnamefont {D.~A.}\ \bibnamefont
  {Huse}},\ }\href@noop {} {\bibfield  {journal} {\bibinfo  {journal} {Physical
  Review B}\ }\textbf {\bibinfo {volume} {24}},\ \bibinfo {pages} {5180}
  (\bibinfo {year} {1981})}\BibitemShut {NoStop}%
\bibitem [{\citenamefont {Haldane}(1983)}]{Haldane:1983ru}%
  \BibitemOpen
  \bibfield  {author} {\bibinfo {author} {\bibfnamefont {F.~D.~M.}\
  \bibnamefont {Haldane}},\ }\href@noop {} {\bibfield  {journal} {\bibinfo
  {journal} {Physical Review Letters}\ }\textbf {\bibinfo {volume} {50}},\
  \bibinfo {pages} {1153} (\bibinfo {year} {1983})}\BibitemShut {NoStop}%
\bibitem [{\citenamefont {Fendley}(2012)}]{Fendley:2012hw}%
  \BibitemOpen
  \bibfield  {author} {\bibinfo {author} {\bibfnamefont {P.}~\bibnamefont
  {Fendley}},\ }\href@noop {} {\bibfield  {journal} {\bibinfo  {journal}
  {Journal of Statistical Mechanics: Theory and Experiment}\ }\textbf {\bibinfo
  {volume} {2012}},\ \bibinfo {pages} {P11020} (\bibinfo {year}
  {2012})}\BibitemShut {NoStop}%
\bibitem [{\citenamefont {Rotteler}\ and\ \citenamefont
  {Wocjan}(2006)}]{Rotteler:2006km}%
  \BibitemOpen
  \bibfield  {author} {\bibinfo {author} {\bibfnamefont {M.}~\bibnamefont
  {Rotteler}}\ and\ \bibinfo {author} {\bibfnamefont {P.}~\bibnamefont
  {Wocjan}},\ }\href@noop {} {\bibfield  {journal} {\bibinfo  {journal} {IEEE
  transactions on information theory}\ } (\bibinfo {year} {2006})}\BibitemShut
  {NoStop}%
\bibitem [{\citenamefont {Stollsteimer}\ and\ \citenamefont
  {Mahler}(2001)}]{PhysRevA.64.052301}%
  \BibitemOpen
  \bibfield  {author} {\bibinfo {author} {\bibfnamefont {M.}~\bibnamefont
  {Stollsteimer}}\ and\ \bibinfo {author} {\bibfnamefont {G.~u.}\ \bibnamefont
  {Mahler}},\ }\href@noop {} {\bibfield  {journal} {\bibinfo  {journal}
  {Physical Review A}\ }\textbf {\bibinfo {volume} {64}},\ \bibinfo {pages}
  {052301} (\bibinfo {year} {2001})}\BibitemShut {NoStop}%
\bibitem [{\citenamefont {Nielsen}\ \emph {et~al.}(2002)\citenamefont
  {Nielsen}, \citenamefont {Bremner}, \citenamefont {Dodd}, \citenamefont
  {Childs},\ and\ \citenamefont {Dawson}}]{PhysRevA.66.022317}%
  \BibitemOpen
  \bibfield  {author} {\bibinfo {author} {\bibfnamefont {M.~A.}\ \bibnamefont
  {Nielsen}}, \bibinfo {author} {\bibfnamefont {M.~J.}\ \bibnamefont
  {Bremner}}, \bibinfo {author} {\bibfnamefont {J.~L.}\ \bibnamefont {Dodd}},
  \bibinfo {author} {\bibfnamefont {A.~M.}\ \bibnamefont {Childs}}, \ and\
  \bibinfo {author} {\bibfnamefont {C.~M.}\ \bibnamefont {Dawson}},\
  }\href@noop {} {\bibfield  {journal} {\bibinfo  {journal} {Physical Review
  A}\ }\textbf {\bibinfo {volume} {66}},\ \bibinfo {pages} {022317} (\bibinfo
  {year} {2002})}\BibitemShut {NoStop}%
\bibitem [{Note1()}]{Note1}%
  \BibitemOpen
  \bibinfo {note} {More specifically, we consider a sequence $\protect
  \{P_i\protect \}$ such that $P_k \protect \dots P_2 P_1 = \protect \mathbb
  {I}$ by appropriately setting either $P_1$ or $P_k$.}\BibitemShut {Stop}%
\bibitem [{\citenamefont {Mori}\ \emph {et~al.}(2016)\citenamefont {Mori},
  \citenamefont {Kuwahara},\ and\ \citenamefont {Saito}}]{Mori:2016wb}%
  \BibitemOpen
  \bibfield  {author} {\bibinfo {author} {\bibfnamefont {T.}~\bibnamefont
  {Mori}}, \bibinfo {author} {\bibfnamefont {T.}~\bibnamefont {Kuwahara}}, \
  and\ \bibinfo {author} {\bibfnamefont {K.}~\bibnamefont {Saito}},\
  }\href@noop {} {\bibfield  {journal} {\bibinfo  {journal} {Physical Review
  Letters}\ }\textbf {\bibinfo {volume} {116}},\ \bibinfo {pages} {120401}
  (\bibinfo {year} {2016})}\BibitemShut {NoStop}%
\bibitem [{\citenamefont {Abanin}\ \emph {et~al.}(2017)\citenamefont {Abanin},
  \citenamefont {De~Roeck}, \citenamefont {Ho},\ and\ \citenamefont
  {Huveneers}}]{Abanin:2015uh}%
  \BibitemOpen
  \bibfield  {author} {\bibinfo {author} {\bibfnamefont {D.~A.}\ \bibnamefont
  {Abanin}}, \bibinfo {author} {\bibfnamefont {W.}~\bibnamefont {De~Roeck}},
  \bibinfo {author} {\bibfnamefont {W.~W.}\ \bibnamefont {Ho}}, \ and\ \bibinfo
  {author} {\bibfnamefont {F.}~\bibnamefont {Huveneers}},\ }\href@noop {}
  {\bibfield  {journal} {\bibinfo  {journal} {Physical Review B}\ }\textbf
  {\bibinfo {volume} {95}},\ \bibinfo {pages} {014112} (\bibinfo {year}
  {2017})}\BibitemShut {NoStop}%
\bibitem [{\citenamefont {Abanin}\ \emph {et~al.}(2015)\citenamefont {Abanin},
  \citenamefont {De~Roeck}, \citenamefont {Huveneers},\ and\ \citenamefont
  {Ho}}]{Abanin:2015uy}%
  \BibitemOpen
  \bibfield  {author} {\bibinfo {author} {\bibfnamefont {D.}~\bibnamefont
  {Abanin}}, \bibinfo {author} {\bibfnamefont {W.}~\bibnamefont {De~Roeck}},
  \bibinfo {author} {\bibfnamefont {F.}~\bibnamefont {Huveneers}}, \ and\
  \bibinfo {author} {\bibfnamefont {W.~W.}\ \bibnamefont {Ho}},\ }\href@noop {}
  {\bibfield  {journal} {\bibinfo  {journal} {arXiv.org}\ } (\bibinfo {year}
  {2015})},\ \Eprint {http://arxiv.org/abs/1509.05386v2} {1509.05386v2}
  \BibitemShut {NoStop}%
\bibitem [{\citenamefont {Kuwahara}\ \emph {et~al.}(2016)\citenamefont
  {Kuwahara}, \citenamefont {Mori},\ and\ \citenamefont
  {Saito}}]{Kuwahara:2016dh}%
  \BibitemOpen
  \bibfield  {author} {\bibinfo {author} {\bibfnamefont {T.}~\bibnamefont
  {Kuwahara}}, \bibinfo {author} {\bibfnamefont {T.}~\bibnamefont {Mori}}, \
  and\ \bibinfo {author} {\bibfnamefont {K.}~\bibnamefont {Saito}},\
  }\href@noop {} {\bibfield  {journal} {\bibinfo  {journal} {Annals of
  Physics}\ }\textbf {\bibinfo {volume} {367}},\ \bibinfo {pages} {96}
  (\bibinfo {year} {2016})}\BibitemShut {NoStop}%
\bibitem [{sup()}]{supp_info}%
  \BibitemOpen
  \href@noop {} {}\bibinfo {note} {See Supplementary Materials for detailed
  information}\BibitemShut {NoStop}%
\bibitem [{\citenamefont {D{\"u}r}\ \emph {et~al.}(2000)\citenamefont
  {D{\"u}r}, \citenamefont {Cirac}, \citenamefont {Lewenstein},\ and\
  \citenamefont {Bru{\ss}}}]{PhysRevA.61.062313}%
  \BibitemOpen
  \bibfield  {author} {\bibinfo {author} {\bibfnamefont {W.}~\bibnamefont
  {D{\"u}r}}, \bibinfo {author} {\bibfnamefont {I.}~\bibnamefont {Cirac}},
  \bibinfo {author} {\bibfnamefont {M.}~\bibnamefont {Lewenstein}}, \ and\
  \bibinfo {author} {\bibfnamefont {D.}~\bibnamefont {Bru{\ss}}},\ }\href@noop
  {} {\bibfield  {journal} {\bibinfo  {journal} {Physical Review A}\ }\textbf
  {\bibinfo {volume} {61}},\ \bibinfo {pages} {062313} (\bibinfo {year}
  {2000})}\BibitemShut {NoStop}%
\bibitem [{Note2()}]{Note2}%
  \BibitemOpen
  \bibinfo {note} {We consider $\protect \mathcal {U}$ to be constructed from a
  set of composite rotations made from simple pulses up to constant depth. This
  is a particularly natural restriction in the context of experiments, where
  finite precision limits the available operations. Also, noting that the
  actual pulse to be applied is $p_i = u_i u_{i-1}^\dagger $, we assume that if
  $u_i$ and $u_{i-1}$ are experimentally feasible, then so is $p_i$. Finally, a
  reduction of the interaction strength is inevitable and captured by $\beta $;
  our algorithm will give the maximum possible value of $\beta $ within the
  given constraints.}\BibitemShut {Stop}%
\bibitem [{\citenamefont {Bertsimas}\ \emph {et~al.}(1997)\citenamefont
  {Bertsimas}, \citenamefont {Tsitsiklis},\ and\ \citenamefont
  {Tsitsiklis}}]{Bertsimas:1997va}%
  \BibitemOpen
  \bibfield  {author} {\bibinfo {author} {\bibfnamefont {D.}~\bibnamefont
  {Bertsimas}}, \bibinfo {author} {\bibfnamefont {J.~N.}\ \bibnamefont
  {Tsitsiklis}}, \ and\ \bibinfo {author} {\bibfnamefont {J.}~\bibnamefont
  {Tsitsiklis}},\ }\href@noop {} {\emph {\bibinfo {title} {{Introduction to
  Linear Optimization (Athena Scientific Series in Optimization and Neural
  Computation, 6)}}}}\ (\bibinfo  {publisher} {Athena Scientific},\ \bibinfo
  {year} {1997})\BibitemShut {NoStop}%
\bibitem [{\citenamefont {Kucsko}\ \emph {et~al.}(2016)\citenamefont {Kucsko},
  \citenamefont {Choi}, \citenamefont {Choi}, \citenamefont {Maurer},
  \citenamefont {Sumiya}, \citenamefont {Onoda}, \citenamefont {Isoya},
  \citenamefont {Jelezko}, \citenamefont {Demler}, \citenamefont {Yao},\ and\
  \citenamefont {Lukin}}]{Kucsko:2016tn}%
  \BibitemOpen
  \bibfield  {author} {\bibinfo {author} {\bibfnamefont {G.}~\bibnamefont
  {Kucsko}}, \bibinfo {author} {\bibfnamefont {S.}~\bibnamefont {Choi}},
  \bibinfo {author} {\bibfnamefont {J.}~\bibnamefont {Choi}}, \bibinfo {author}
  {\bibfnamefont {P.~C.}\ \bibnamefont {Maurer}}, \bibinfo {author}
  {\bibfnamefont {H.}~\bibnamefont {Sumiya}}, \bibinfo {author} {\bibfnamefont
  {S.}~\bibnamefont {Onoda}}, \bibinfo {author} {\bibfnamefont
  {J.}~\bibnamefont {Isoya}}, \bibinfo {author} {\bibfnamefont
  {F.}~\bibnamefont {Jelezko}}, \bibinfo {author} {\bibfnamefont
  {E.}~\bibnamefont {Demler}}, \bibinfo {author} {\bibfnamefont {N.~Y.}\
  \bibnamefont {Yao}}, \ and\ \bibinfo {author} {\bibfnamefont {M.~D.}\
  \bibnamefont {Lukin}},\ }\href@noop {} {\  (\bibinfo {year} {2016})},\
  \Eprint {http://arxiv.org/abs/1609.08216} {1609.08216} \BibitemShut {NoStop}%
\bibitem [{\citenamefont {Prakash}\ \emph {et~al.}(2016)\citenamefont
  {Prakash}, \citenamefont {West},\ and\ \citenamefont {Wei}}]{Prakash:2016df}%
  \BibitemOpen
  \bibfield  {author} {\bibinfo {author} {\bibfnamefont {A.}~\bibnamefont
  {Prakash}}, \bibinfo {author} {\bibfnamefont {C.~G.}\ \bibnamefont {West}}, \
  and\ \bibinfo {author} {\bibfnamefont {T.~C.}\ \bibnamefont {Wei}},\
  }\href@noop {} {\bibfield  {journal} {\bibinfo  {journal} {Physical Review
  B}\ }\textbf {\bibinfo {volume} {94}},\ \bibinfo {pages} {045136} (\bibinfo
  {year} {2016})}\BibitemShut {NoStop}%
\bibitem [{\citenamefont {Affleck}\ \emph {et~al.}(1987)\citenamefont
  {Affleck}, \citenamefont {Kennedy}, \citenamefont {Lieb},\ and\ \citenamefont
  {Tasaki}}]{Affleck:1987jy}%
  \BibitemOpen
  \bibfield  {author} {\bibinfo {author} {\bibfnamefont {I.}~\bibnamefont
  {Affleck}}, \bibinfo {author} {\bibfnamefont {T.}~\bibnamefont {Kennedy}},
  \bibinfo {author} {\bibfnamefont {E.~H.}\ \bibnamefont {Lieb}}, \ and\
  \bibinfo {author} {\bibfnamefont {H.}~\bibnamefont {Tasaki}},\ }\href@noop {}
  {\bibfield  {journal} {\bibinfo  {journal} {Physical Review Letters}\
  }\textbf {\bibinfo {volume} {59}},\ \bibinfo {pages} {799} (\bibinfo {year}
  {1987})}\BibitemShut {NoStop}%
\bibitem [{\citenamefont {D'Alessio}\ and\ \citenamefont
  {Rigol}(2014)}]{DAlessio:2014fg}%
  \BibitemOpen
  \bibfield  {author} {\bibinfo {author} {\bibfnamefont {L.}~\bibnamefont
  {D'Alessio}}\ and\ \bibinfo {author} {\bibfnamefont {M.}~\bibnamefont
  {Rigol}},\ }\href@noop {} {\bibfield  {journal} {\bibinfo  {journal}
  {Physical Review X}\ }\textbf {\bibinfo {volume} {4}},\ \bibinfo {pages}
  {041048} (\bibinfo {year} {2014})}\BibitemShut {NoStop}%
\bibitem [{\citenamefont {Lazarides}\ \emph {et~al.}(2014)\citenamefont
  {Lazarides}, \citenamefont {Das},\ and\ \citenamefont
  {Moessner}}]{Lazarides:2014ie}%
  \BibitemOpen
  \bibfield  {author} {\bibinfo {author} {\bibfnamefont {A.}~\bibnamefont
  {Lazarides}}, \bibinfo {author} {\bibfnamefont {A.}~\bibnamefont {Das}}, \
  and\ \bibinfo {author} {\bibfnamefont {R.}~\bibnamefont {Moessner}},\
  }\href@noop {} {\bibfield  {journal} {\bibinfo  {journal} {Physical Review
  E}\ }\textbf {\bibinfo {volume} {90}},\ \bibinfo {pages} {012110} (\bibinfo
  {year} {2014})}\BibitemShut {NoStop}%
\bibitem [{\citenamefont {Ponte}\ \emph {et~al.}(2015)\citenamefont {Ponte},
  \citenamefont {Chandran}, \citenamefont {Papi{\'c}},\ and\ \citenamefont
  {Abanin}}]{Ponte:2015vj}%
  \BibitemOpen
  \bibfield  {author} {\bibinfo {author} {\bibfnamefont {P.}~\bibnamefont
  {Ponte}}, \bibinfo {author} {\bibfnamefont {A.}~\bibnamefont {Chandran}},
  \bibinfo {author} {\bibfnamefont {Z.}~\bibnamefont {Papi{\'c}}}, \ and\
  \bibinfo {author} {\bibfnamefont {D.~A.}\ \bibnamefont {Abanin}},\
  }\href@noop {} {\bibfield  {journal} {\bibinfo  {journal} {Annals of
  Physics}\ } (\bibinfo {year} {2015})}\BibitemShut {NoStop}%
\bibitem [{\citenamefont {Choi}\ \emph
  {et~al.}(2017{\natexlab{b}})\citenamefont {Choi}, \citenamefont {Abanin},\
  and\ \citenamefont {Lukin}}]{DIMBLChoi:2017wm}%
  \BibitemOpen
  \bibfield  {author} {\bibinfo {author} {\bibfnamefont {S.}~\bibnamefont
  {Choi}}, \bibinfo {author} {\bibfnamefont {D.~A.}\ \bibnamefont {Abanin}}, \
  and\ \bibinfo {author} {\bibfnamefont {M.~D.}\ \bibnamefont {Lukin}},\
  }\href@noop {} {\bibfield  {journal} {\bibinfo  {journal} {arXiv.org}\ }
  (\bibinfo {year} {2017}{\natexlab{b}})},\ \Eprint
  {http://arxiv.org/abs/1703.03809v1} {1703.03809v1} \BibitemShut {NoStop}%
\bibitem [{\citenamefont {Choi}\ \emph {et~al.}(2015)\citenamefont {Choi},
  \citenamefont {Yao}, \citenamefont {Gopalakrishnan},\ and\ \citenamefont
  {Lukin}}]{QCMBLChoi:2015}%
  \BibitemOpen
  \bibfield  {author} {\bibinfo {author} {\bibfnamefont {S.}~\bibnamefont
  {Choi}}, \bibinfo {author} {\bibfnamefont {N.~Y.}\ \bibnamefont {Yao}},
  \bibinfo {author} {\bibfnamefont {S.}~\bibnamefont {Gopalakrishnan}}, \ and\
  \bibinfo {author} {\bibfnamefont {M.~D.}\ \bibnamefont {Lukin}},\ }\href@noop
  {} {\bibfield  {journal} {\bibinfo  {journal} {arXiv.org}\ } (\bibinfo {year}
  {2015})},\ \Eprint {http://arxiv.org/abs/1508.06992v1} {1508.06992v1}
  \BibitemShut {NoStop}%
\bibitem [{\citenamefont {Yao}\ \emph {et~al.}(2015)\citenamefont {Yao},
  \citenamefont {Laumann},\ and\ \citenamefont
  {Vishwanath}}]{MBLStateTransferYao:2015}%
  \BibitemOpen
  \bibfield  {author} {\bibinfo {author} {\bibfnamefont {N.~Y.}\ \bibnamefont
  {Yao}}, \bibinfo {author} {\bibfnamefont {C.~R.}\ \bibnamefont {Laumann}}, \
  and\ \bibinfo {author} {\bibfnamefont {A.}~\bibnamefont {Vishwanath}},\
  }\href@noop {} {\bibfield  {journal} {\bibinfo  {journal} {arXiv.org}\ }
  (\bibinfo {year} {2015})},\ \Eprint {http://arxiv.org/abs/1508.06995v1}
  {1508.06995v1} \BibitemShut {NoStop}%
\end{thebibliography}%


\begin{thebibliography}{7}%
\makeatletter
\providecommand \@ifxundefined [1]{%
 \@ifx{#1\undefined}
}%
\providecommand \@ifnum [1]{%
 \ifnum #1\expandafter \@firstoftwo
 \else \expandafter \@secondoftwo
 \fi
}%
\providecommand \@ifx [1]{%
 \ifx #1\expandafter \@firstoftwo
 \else \expandafter \@secondoftwo
 \fi
}%
\providecommand \natexlab [1]{#1}%
\providecommand \enquote  [1]{``#1''}%
\providecommand \bibnamefont  [1]{#1}%
\providecommand \bibfnamefont [1]{#1}%
\providecommand \citenamefont [1]{#1}%
\providecommand \href@noop [0]{\@secondoftwo}%
\providecommand \href [0]{\begingroup \@sanitize@url \@href}%
\providecommand \@href[1]{\@@startlink{#1}\@@href}%
\providecommand \@@href[1]{\endgroup#1\@@endlink}%
\providecommand \@sanitize@url [0]{\catcode `\\12\catcode `\$12\catcode
  `\&12\catcode `\#12\catcode `\^12\catcode `\_12\catcode `\%12\relax}%
\providecommand \@@startlink[1]{}%
\providecommand \@@endlink[0]{}%
\providecommand \url  [0]{\begingroup\@sanitize@url \@url }%
\providecommand \@url [1]{\endgroup\@href {#1}{\urlprefix }}%
\providecommand \urlprefix  [0]{URL }%
\providecommand \Eprint [0]{\href }%
\providecommand \doibase [0]{http://dx.doi.org/}%
\providecommand \selectlanguage [0]{\@gobble}%
\providecommand \bibinfo  [0]{\@secondoftwo}%
\providecommand \bibfield  [0]{\@secondoftwo}%
\providecommand \translation [1]{[#1]}%
\providecommand \BibitemOpen [0]{}%
\providecommand \bibitemStop [0]{}%
\providecommand \bibitemNoStop [0]{.\EOS\space}%
\providecommand \EOS [0]{\spacefactor3000\relax}%
\providecommand \BibitemShut  [1]{\csname bibitem#1\endcsname}%
\let\auto@bib@innerbib\@empty
%</preamble>
\bibitem [{\citenamefont {Waugh}\ \emph {et~al.}(1968)\citenamefont {Waugh},
  \citenamefont {Huber},\ and\ \citenamefont {Haeberlen}}]{Waugh:1968im}%
  \BibitemOpen
  \bibfield  {author} {\bibinfo {author} {\bibfnamefont {J.~S.}\ \bibnamefont
  {Waugh}}, \bibinfo {author} {\bibfnamefont {L.~M.}\ \bibnamefont {Huber}}, \
  and\ \bibinfo {author} {\bibfnamefont {U.}~\bibnamefont {Haeberlen}},\
  }\href@noop {} {\bibfield  {journal} {\bibinfo  {journal} {Physical Review
  Letters}\ }\textbf {\bibinfo {volume} {20}},\ \bibinfo {pages} {180}
  (\bibinfo {year} {1968})}\BibitemShut {NoStop}%
\bibitem [{\citenamefont {Chen}\ \emph
  {et~al.}(2011{\natexlab{a}})\citenamefont {Chen}, \citenamefont {Gu},\ and\
  \citenamefont {Wen}}]{Chen:2011vg}%
  \BibitemOpen
  \bibfield  {author} {\bibinfo {author} {\bibfnamefont {X.}~\bibnamefont
  {Chen}}, \bibinfo {author} {\bibfnamefont {Z.~C.}\ \bibnamefont {Gu}}, \ and\
  \bibinfo {author} {\bibfnamefont {X.~G.}\ \bibnamefont {Wen}},\ }\href@noop
  {} {\bibfield  {journal} {\bibinfo  {journal} {Physical Review B}\ }\textbf
  {\bibinfo {volume} {83}},\ \bibinfo {pages} {035107} (\bibinfo {year}
  {2011}{\natexlab{a}})}\BibitemShut {NoStop}%
\bibitem [{\citenamefont {Chen}\ \emph
  {et~al.}(2011{\natexlab{b}})\citenamefont {Chen}, \citenamefont {Gu},\ and\
  \citenamefont {Wen}}]{Chen:2011fp}%
  \BibitemOpen
  \bibfield  {author} {\bibinfo {author} {\bibfnamefont {X.}~\bibnamefont
  {Chen}}, \bibinfo {author} {\bibfnamefont {Z.-C.}\ \bibnamefont {Gu}}, \ and\
  \bibinfo {author} {\bibfnamefont {X.-G.}\ \bibnamefont {Wen}},\ }\href@noop
  {} {\bibfield  {journal} {\bibinfo  {journal} {Physical Review B}\ }\textbf
  {\bibinfo {volume} {84}},\ \bibinfo {pages} {235128} (\bibinfo {year}
  {2011}{\natexlab{b}})}\BibitemShut {NoStop}%
\bibitem [{\citenamefont {Schuch}\ \emph {et~al.}(2011)\citenamefont {Schuch},
  \citenamefont {Perez-Garcia},\ and\ \citenamefont {Cirac}}]{Schuch:2011gm}%
  \BibitemOpen
  \bibfield  {author} {\bibinfo {author} {\bibfnamefont {N.}~\bibnamefont
  {Schuch}}, \bibinfo {author} {\bibfnamefont {D.}~\bibnamefont
  {Perez-Garcia}}, \ and\ \bibinfo {author} {\bibfnamefont {I.}~\bibnamefont
  {Cirac}},\ }\href@noop {} {\bibfield  {journal} {\bibinfo  {journal}
  {Physical Review B}\ }\textbf {\bibinfo {volume} {84}},\ \bibinfo {pages}
  {165139} (\bibinfo {year} {2011})}\BibitemShut {NoStop}%
\bibitem [{\citenamefont {Pollmann}\ and\ \citenamefont
  {Turner}(2012)}]{Pollmann:2012uj}%
  \BibitemOpen
  \bibfield  {author} {\bibinfo {author} {\bibfnamefont {F.}~\bibnamefont
  {Pollmann}}\ and\ \bibinfo {author} {\bibfnamefont {A.~M.}\ \bibnamefont
  {Turner}},\ }\href@noop {} {\bibfield  {journal} {\bibinfo  {journal}
  {Physical Review B}\ } (\bibinfo {year} {2012})}\BibitemShut {NoStop}%
\bibitem [{\citenamefont {Prakash}\ \emph {et~al.}(2016)\citenamefont
  {Prakash}, \citenamefont {West},\ and\ \citenamefont {Wei}}]{Prakash:2016df}%
  \BibitemOpen
  \bibfield  {author} {\bibinfo {author} {\bibfnamefont {A.}~\bibnamefont
  {Prakash}}, \bibinfo {author} {\bibfnamefont {C.~G.}\ \bibnamefont {West}}, \
  and\ \bibinfo {author} {\bibfnamefont {T.~C.}\ \bibnamefont {Wei}},\
  }\href@noop {} {\bibfield  {journal} {\bibinfo  {journal} {Physical Review
  B}\ }\textbf {\bibinfo {volume} {94}},\ \bibinfo {pages} {045136} (\bibinfo
  {year} {2016})}\BibitemShut {NoStop}%
\bibitem [{\citenamefont {Vidal}(2007)}]{Vidal:2007hx}%
  \BibitemOpen
  \bibfield  {author} {\bibinfo {author} {\bibfnamefont {G.}~\bibnamefont
  {Vidal}},\ }\href@noop {} {\bibfield  {journal} {\bibinfo  {journal}
  {Physical Review Letters}\ }\textbf {\bibinfo {volume} {98}},\ \bibinfo
  {pages} {070201} (\bibinfo {year} {2007})}\BibitemShut {NoStop}%
\end{thebibliography}%
\end{document}

\title{Supplementary Materials for Dynamical engineering of interactions in qudit ensembles} 

\author{Soonwon Choi}
\affiliation{Department of Physics, Harvard University, Cambridge, Massachusetts 02138, USA}

\author{Norman Y. Yao}
\affiliation{Department of Physics, University of California Berkeley, Berkeley, California 94720, USA}

\author{Mikhail D. Lukin}
\affiliation{Department of Physics, Harvard University, Cambridge, Massachusetts 02138, USA}

\maketitle

\section{Generalized Gell-Mann matrices}

In the main text, we parametrize interactions using a set of trace orthonormal matrices $\{\lambda_\mu\}$.
Here, for completeness, we present explicit expressions of $\{\lambda_\mu\}$ for $d=2$ and $d=3$.
For $d>3$, we provide a general method for constructing $\{\lambda_\mu\}$.

For $d=2$ (qubits) the operator basis $\{\lambda_\mu \}$ coincides with Pauli matrices:
\begin{align}
    \lambda_1 = \sigma^x = 
    \left(
    \begin{array}{cc}
    0 & 1 \\
    1 & 0
    \end{array}
    \right),\;\;\;
    \lambda_2 = \sigma^y = 
    \left(
    \begin{array}{cc}
    0 & -i \\
    i & 0
    \end{array}
    \right),\;\;\;
    \lambda_3 = \sigma^z = 
    \left(
    \begin{array}{cc}
    1 & 0 \\
    0 & -1
    \end{array}
    \right).
\end{align}
As required, these matrices satisfy the trace orthonormality $\trace{\lambda_\mu \lambda_\nu} = 2\delta_{\mu \nu}$ and, together with identity $\mathbb{1}_2$, form a basis for two dimensional Hermitian matrices.
For $d=3$ (spin-1 particles or qutrits), we choose $\{\lambda_\mu\}$ as so-called Gell-Mann matrices:
\begin{align}
    \lambda_1 &=\left(
\begin{array}{ccc}
 0 & 1 & 0 \\
 1 & 0 & 0 \\
 0 & 0 & 0 
\end{array}
\right),
&\lambda_2 &= \left(
\begin{array}{ccc}
 0 & 0 & 0 \\
 0 & 0 & 1 \\
 0 & 1 & 0 
\end{array}
\right),
&\lambda_3 & = \left(
\begin{array}{ccc}
 0 & 0 & 1 \\
 0 & 0 & 0 \\
 1 & 0 & 0 
\end{array}
\right),\\
\lambda_4 &= \left(
\begin{array}{ccc}
 0 & -i & 0 \\
 i & 0 & 0 \\
 0 & 0 & 0 
\end{array}
\right), 
& \lambda_5 &= \left(
\begin{array}{ccc}
 0 & 0 & 0 \\
 0 & 0 & -i \\
 0 & i & 0 
\end{array}
\right),
& \lambda_6 &= \left(
\begin{array}{ccc}
 0 & 0 & -i \\
 0 & 0 & 0 \\
 i & 0 & 0 
\end{array}
\right),\\
\lambda_7 &= \left(
\begin{array}{ccc}
 1 & 0 & 0 \\
 0 & -1 & 0 \\
 0 & 0 & 0 
\end{array}
\right),
& \lambda_8 &= \frac{1}{\sqrt{3}} \left(
\begin{array}{ccc}
 1 & 0 & 0 \\
 0 & 1 & 0 \\
 0 & 0 & -2
\end{array}
\right).
\end{align}
Again, these matrices are traceless and orthonormal (normalized to $\trace{\lambda_\mu \lambda_\mu}=2$), and form a basis for three dimensional Hermitian matrices together with the identity $\mathbb{1}_3$.
Note that three matrices in the first and the second lines are purely real and imaginary, respectively, and the last two matrices are real and diagonal.
For a generic $d$, we construct $m=d^2-1$ matrices in the following way. Let $E_{ij}$ be a matrix with an element $1$ at the $i$-th row and $j$-th column and zeros elsewhere. We define first $d(d-1)/2$ matrices as
\begin{align}
    \lambda_\mu = E_{i_\mu, j_\mu} + E_{j_\mu, i_\mu} \;\; \textrm{ for } \mu \in \{1, 2, \dots , d(d-1)/2\},
\end{align}
where $(i_\mu, j_\mu)$ enumerates all possible $d(d-1)/2$ combinations of $i < j$ pairs.
The next $d(d-1)/2$ matrices are similarly defined as
\begin{align}
    \lambda_\mu = -i E_{i_\mu, j_\mu} + i E_{j_\mu, i_\mu} \;\; \textrm{ for } \mu \in \{d(d-1)/2+1,\cdots,  d(d-1)\}.
\end{align}
Finally, the remaining $d-1$ matrices are real, diagonal, and defined as 
\begin{align}
    \lambda_\mu =\frac{1}{\sqrt{k_\mu(k_\mu-1)/2}} \left(\sum_{i=1}^{k_\mu-1} E_{ii} - (k_\mu-1) E_{k_\mu, k_\mu}\right)
    \;\; \textrm{ for } \mu \in \{d(d-1)+1, \dots, d^2-1 \},
\end{align}
where $k_\mu$ enumerates $\{2,3, \cdots, d\}$.
These matrices are traceless by constructions, and their orthonormality can be checked by explicit computations.

\section{Dynamical decoupling of dipolar interactions among spin-1 particles}
In this section, we provide the details of dipolar interactions among spin-1 particles and their decoupling by using our $6$-pulse sequence.
We start with a generic Hamiltonian of the form $H = \sum_i H^{(1)}_i + \sum_{ij} H^{\textrm{d-d}}_{ij}$, where $H_i^{(1)}$ is a single spin Hamiltonian for a particle $i$ and $H_{ij}^{\textrm{d-d}}$ is a pairwise dipolar interaction for a particle pair $i$ and $j$ 
\begin{align}
        H^{\textrm{d-d}}_{ij} =& - \frac{J_0}{r_{ij}^3} \left( 3 \left(\vec{S}_i \cdot \vec{r}_{ij}\right)\left(\vec{S}_j \cdot \vec{r}_{ij}\right)/r_{ij}^2 - \vec{S}_i \cdot \vec{S}_j \right)
\end{align}
with the interaction strength $J_0$, the relative position of the pair $\vec{r}_{ij}$, and the spin-1 vector operators $\vec{S}_i = \left( S^x_i, S^y_i, S^z_i \right)$.
In the absence of single particle terms, the dipolar interactions can be efficiently suppressed by a sequence of $SO(3)$ spin rotations.
This can be understood by rewriting the interactions as $H_{ij}^\textrm{d-d} = -(J_0/r_{ij}^3) \sum_{\mu \nu} S_i^\mu T_{\mu \nu} S_j^\nu$ with a rank-2 tensor
\begin{align}
    T = \left(
    \begin{array}{ccc}
    -1 & 0 & 0\\
    0 & -1 & 0\\
    0 & 0 & 2        
    \end{array}    
    \right),
\end{align}
where we have chosen $\hat{z} \equiv \vec{r}_{ij}/r_{ij}$ without the loss of generality.
As well-known in the nuclear magnetic resonance (NMR) community, the tensor $T$ can be symmetrized to zero upon three $SO(3)$ rotations, effectively decoupling the interaction
~\cite{Waugh:1968im}.

In contrast, under the presence of strong single particle terms, the form of interactions can be effectively modified, making it impossible to decouple them only using $SO(3)$ rotations.
More specifically, we consider strongly anharmonic energy levels of spin-1 particles characterized by a Hamiltonian $H_i^{(1)} = h S_i^z + \Delta (S_i^z)^2 $,
where the first term typically arises from Zeeman coupling to external magnetic field and the second term naturally occurs when spin symmetries are broken, e.g., by quadrupolar couplings for nuclear spins or by spin-orbit couplings for nitrogen vacancy color centers (NV) in diamond. 
In the limit of strong anharmonicity $|h|, |\Delta|, |h\pm\Delta| \gg J_0 /r_{ij}^3$, as satisfied by most of experiments with solid state NMR or high density NV ensembles, the conservation of energy suppresses some of spin exchange processes in $H_{ij}^\textrm{d-d}$.
The resultant effective interactions can be obtained in the interacting picture with a transformation 
\begin{align}
\label{eqn:dipole_eff}
H_\textrm{eff} (t) = U_0^{-1}(t) H U_0(t) - i U_0^{-1} (t) \frac{\partial}{\partial t} U_0(t),
\end{align}
 where $U_0(t) = \exp{\left[ -i \left( \sum_i h S_i^z + \Delta (S_i^z)^2 \right) t\right]}$.
Ignoring energy non-conserving terms (secular approximations), the effective interactions become
$H_\textrm{eff} \approx \sum_{ij}  \frac{J_0}{r^3} ( 1- 3 \cos^2 \theta) \sum_{\mu \nu} C^\textrm{eff}_{\mu \nu}\; \lambda_\mu \otimes \lambda_\nu$
with $\cos \theta = \hat{z} \cdot \vec{r} / |\vec{r}|$ and 
\begin{align}
    C^\textrm{eff} = -\frac{1}{4} \left(
\begin{array}{cccccccc}
1 & 0 & 0 & 0 & 0 & 0 & 0 & 0\\
0 & 0 & 0 & 0 & 0 & 0 & 0 & 0\\
0 & 0 & 1 & 0 & 0 & 0 & 0 & 0\\
0 & 0 & 0 & 1 & 0 & 0 & 0 & 0\\
0 & 0 & 0 & 0 & 0 & 0 & 0 & 0\\
0 & 0 & 0 & 0 & 0 & 1 & 0 & 0\\
0 & 0 & 0 & 0 & 0 & 0 & -1 & -\sqrt{3}\\
0 & 0 & 0 & 0 & 0 & 0 & -\sqrt{3} & -3\\
\end{array}
 \right).
\end{align}
Identifying $J_{ij} \equiv -(J_0/r^3_{ij})(1-3\cos^2\theta)$ reduces $H_\textrm{eff}$ to the expression given in the main text.
We note that $C^\textrm{eff}$ is traceless and hence allows a complete suppression by a pulse sequence.

In order to find a decoupling pulse sequence, we use our algorithm presented in the main text. We assume that the set of available unitaries $\mathcal{U}$ is limited to composite pulses made out of up to four $\pm\pi$ and $\pm(\pi/2)$-pulses: we define a set of elementary operations $\mathcal{E} = \{\mathbb{1}_3, e^{\pm i \frac{\pi}{2}X_a} , e^{\pm i \pi X_a}, e^{\pm i \frac{\pi}{2}Y_a}, e^{\pm i \pi Y_a}\}$ with $X_a = \lambda_a/2$ and $Y_a = (\lambda_{a+3})/2$ ($a\in \{1, 2, 3\}$) and construct composite pulses $\mathcal{U} = \{u = x_1 x_2 x_3 x_4| x_1, x_2, x_3, x_4 \in \mathcal{E} \}$.
Using a linear programming routine built-in \verb|Mathematica|, we find a set of 6 unitary rotations that average $C^\textrm{eff}$ to zero:
\begin{align}
    u_1& = e^{-i\frac{\pi}{2}X_3}
          e^{-i\frac{\pi}{2}Y_3}
          e^{-i\pi X_1}
          e^{i\frac{\pi}{2}Y_3}
          e^{i\frac{\pi}{2}X_3},
    &u_2& = e^{-i\frac{\pi}{2}X_3}
          e^{-i\frac{\pi}{2}Y_3}
          e^{-i\pi Y_2}
          e^{i\frac{\pi}{2}Y_3}
          e^{i\frac{\pi}{2}X_3},
    &u_3 &=e^{-i\frac{\pi}{2}Y_3}
          e^{i\frac{\pi}{2}X_3}
          e^{i\frac{\pi}{2}Y_3}
          e^{i\frac{\pi}{2}X_3},\\
    u_4& = e^{-i\frac{\pi}{2}Y_3}
          e^{i\frac{\pi}{2}X_3}
          e^{-i\pi X_1}
          e^{i\frac{\pi}{2}Y_3}
          e^{i\frac{\pi}{2}X_3},
    &u_5& = e^{-i\frac{\pi}{2}Y_3}
          e^{i\frac{\pi}{2}X_3}
          e^{-i\pi Y_2}
          e^{i\frac{\pi}{2}Y_3}
          e^{i\frac{\pi}{2}X_3},
    &u_6 &= \mathbb{1}_3.
\end{align}
The corresponding pulse sequence $p_i = u_i u_{i-1}^{\dagger}$ is given as
\begin{align}
    p_1 & = e^{-i\frac{\pi}{2}X_3}
          e^{-i\frac{\pi}{2}Y_3}
          e^{-i\pi X_1}
          e^{i\frac{\pi}{2}Y_3}
          e^{i\frac{\pi}{2}X_3},\\
   p_2& = e^{-i\frac{\pi}{2}X_3}
          e^{-i\frac{\pi}{2}Y_3}
          e^{-i\pi Y_2}
          e^{i\pi X_1}
          e^{i\frac{\pi}{2}Y_3}
          e^{i\frac{\pi}{2}X_3},\\
   p_3 &=e^{-i\frac{\pi}{2}Y_3}
          e^{i\frac{\pi}{2}X_3}
          e^{i\pi Y_2}
          e^{i\frac{\pi}{2}Y_3}
          e^{i\frac{\pi}{2}X_3},\\
   p_4& = e^{-i\frac{\pi}{2}Y_3}
          e^{i\frac{\pi}{2}X_3}
          e^{-i\pi X_1}
          e^{-i\frac{\pi}{2}X_3}
          e^{i\frac{\pi}{2}Y_3},\\
   p_5& = e^{-i\frac{\pi}{2}Y_3}
          e^{i\frac{\pi}{2}X_3}
          e^{-i\pi Y_2}
          e^{i\pi X_1}
          e^{-i\frac{\pi}{2}X_3}
          e^{i\frac{\pi}{2}Y_3},\\
   p_6 &= e^{-i\frac{\pi}{2}X_3}
          e^{-i\frac{\pi}{2}Y_3}
          e^{i\pi Y_2}
          e^{-i\frac{\pi}{2}X_3}
          e^{i\frac{\pi}{2}Y_3}.          
\end{align}
The numerical simulation presented in the main text is based on the exact diagonalization of the time evolution over one period $U_T = P_6e^{-iH_d T/6} P_5 \dots P_1e^{-iH_d T/6}$, where $P_i \equiv p_i^{\otimes N}$ for $N=6$ particles.

We note that the order of $u_i$ is not important within our approximations.
Therefore, by rearranging the order of $u_i$, one can significantly simplify the corresponding pulse sequence $p'_i$.
Also, once a composite pulse $p'_i$ is identified as a sequence of elementary operations, one can further ``compress'' it using algebraic identities of $SU(d)$ group.
For instance, the above dynamical decoupling can be also achieved via the following sequence
\begin{align}
    p'_1 &= e^{i\pi \frac{X_1+X_2}{\sqrt{2}} }\\
    p'_2 &= e^{-i\pi \frac{Y_1+X_2}{\sqrt{2}} }\\
    p'_3 &= e^{i\frac{\pi}{2} Y_3}e^{i\frac{\pi}{4}S_z}\\
    p'_4 &= e^{i\pi \frac{X_1+Y_2}{\sqrt{2}} }\\
    p'_5 &= e^{-i\pi \frac{X_1-Y_2}{\sqrt{2}} }\\
    p'_6 & = (p'_3)^{-1}.
\end{align} 
In experiments with NMR or high density NV ensembles, the pulses $p'_1$, $p'_2$, $p'_4$, and $p'_5$ can be implemented by simultaneous microwave driving of two transitions with appropriate phase choices
while $p'_3$ and $p'_6$ can be realized by using AC stark shifts and two-photon Raman transition. Also, $p'_3$ and $p'_6$ can be decomposed to four short pulses as provided in the figure in the main text.

\section{Symmetrizing a pulse sequence}
In this work we approximate the effective Hamiltonian by truncating Magnus expansion in the zeroth order.
However, given any pulse sequence one can always improve it such that the effective Hamiltonian is also correct up to the first order. For a pulse sequence $\{P_1, \dots P_k\}$ followed by free evolutions $\{\tau_1, \dots, \tau_k\}$, the first order correction in Magnus expansion is given by 
\begin{align}
\label{eqn:magnus_first_order}
    H_\textrm{eff}^{(1)} = -\frac{i}{2T} \sum_{i>j}   [\tau_i \bar{H}_i, \tau_j\bar{H}_j],
\end{align}
where $\bar{H}_i = U_i^\dagger H U_i$ with $U_i = P_i P_{i-1} \dots P_1$ as defined in the main text. The key idea is to appropriately \emph{symmetrize} a pulse sequence such that $H_\textrm{eff}^{(1)}$ exactly vanishes.
More specifically, we now consider a modified pulse sequence of total period $2T$, where the evolution in the first half remains the same while the pulses in the second half is time reversed:
\begin{align}
    U_\textrm{sym}(2T) =P_1^{-1}e^{-iH\tau_1}P_2^{-1}e^{-iH\tau_2}\dots P_k^{-1} e^{-iH\tau_k} \;e^{-iH\tau_k}P_k \dots e^{-iH\tau_2}P_2  e^{-iH\tau_1}P_1.
\end{align}
In the toggling frame, $U_\textrm{sym}(2T)$ can be written as
\begin{align}
    U_\textrm{sym}(2T) =
    e^{-i\bar{H}_1 \tau_1}
    e^{-i\bar{H}_2 \tau_2}
    \dots
    e^{-i\bar{H}_k \tau_k}
    e^{-i\bar{H}_k \tau_k}
    \dots
    e^{-i\bar{H}_2 \tau_2}
    e^{-i\bar{H}_1 \tau_1},
\end{align}
where one finds that every $\bar{H}_i$ appears twice, each in the first (A) and the second (B) half of the period in reversed orders.
For convenience let us denote the pair of $\bar{H}_i$ as $\bar{H}_i^A$ and $\bar{H}_i^B$ depending on their positions. Now the first order correction becomes 
\begin{align}
    H_\textrm{eff}^{(1)} 
    &= -\frac{i}{4T}
    \left( 
    \sum_{i>j \in A} [\tau_i \bar{H}_i^A, \tau_j\bar{H}_j^A] +  
    \sum_{i<j \in B} [\tau_i \bar{H}_i^B, \tau_j\bar{H}_j^B] +
    \sum_{i \in B, j \in A} [\tau_i \bar{H}_i^B, \tau_j\bar{H}_j^A]
    \right)\\
    &= -\frac{i}{4T}
    \left( 
     \left[ \sum_{i\in B}\tau_i \bar{H}_i^B,\sum_{j \in A} \tau_j\bar{H}_j^A\right]\right) = 0,
\end{align}
where the cancellation in the second line is due to the reversed order of indices in B.
Note that, in practice, the last pulse, $P_1^{-1}$, in a symmetrized sequence is immediately compensated by the first pulse, $P_1$, from the next period. Therefore, the number of pulses is generally $2(k-1)$ while the total time duration is exactly doubled to $2T$.

\section{Phase diagram of $H(p,q)$}

The classification of symmetry-protected topological (SPT) phases for bosonic one dimensional systems has been extensively studied \cite{Chen:2011vg,Chen:2011fp,Schuch:2011gm}, and their detections based on numerical methods are  also well known \cite{Pollmann:2012uj,Prakash:2016df}. In particular, we note that Prakash \emph{et al} in Ref.~\cite{Prakash:2016df} investigate the phases of Hamiltonians $H'(r,q)$ that are closely related to our model $H(p,q)$:
\begin{align}
H'(r,q) = H_1 + r H'_{2} + q H_3,
\end{align}
where $H_1$ and $H_3$ are the same as in our case and 
\begin{align}
H'_{2} = \sum_i (S_i^x S_{i+1}^x)^2 + (S_i^y S_{i+1}^y)^2 + (S_i^z S_{i+1}^z)^2.
\end{align}
When $r=0$ this model coincides with our case with $p=0$.
In Ref.~\cite{Prakash:2016df}, Prakash \emph{et al} predicts that $H'(r=0,q)$ with $|q|<1/2$ belongs to a topologically non-trivial phase (the phase C in their Fig.~1) that is equivalent to the AKLT phase.
Also, they show that there are two adjacent, distinct topological phases for $r>0$ (the phases B and D in their Fig.~1). The ground states of these phases respect all symmetries, namely $A_4$ spin rotations, lattice translations, and inversion,
but they are distinguishable by $U(1)$ phases that their wavefunctions  acquire upon the action of a 120$^\circ$ rotation $a \in A_4$.  More specifically, we consider a translationally invariant infinite size matrix product state  $\Gamma^i_{ab}$ for each ground state (with a physical index $i\in\{0,\pm1\}$ and bond indices  $a,b\in\{1,\dots D\}$ ), and study its transformation under the action of internal symmetry group elements $g\in A_4$
\begin{align}
\sum_j u(g)_{ij} \Gamma^j_{ab} =\sum_{a'b'} \chi(g) V^{-1}(g)_{aa'} \Gamma^i_{a'b'} V(g)_{b'b},
\end{align}
where $u(g)_{ij}$ is the unitary representation of a local spin rotation by $g\in A_4$, $\chi(g) \in U(1)$ is an overall phase factor that a wavefunction acquires, and $V(g)$ is a projective representation of $A_4$ with a complex phase $\omega$, i.e., $V(g_1)V(g_2) = \omega(g_1,g_2) V(g_1g_2)$.
When $\omega$ is non-trivial, the corresponding phase is topologically non-trivial as in the case of phases B, C, and D in Ref.~\cite{Prakash:2016df}. The three phases are, however, distinguished by $\chi(g)$; while the the phase C has $\chi(a) = 1$, phases B and D have $\chi(a) = e^{\pm i2 \pi/3}$. 

In our case, the phase diagram looks different from Fig.~1 in Ref.~\cite{Prakash:2016df} owing to a different parametrization of Hamiltonians. Nevertheless, its qualitative features remain similar, and the phase diagram exhibits three SPT phases I, II, and III adjacent to one another. We identify the phases I, II, and III with the phases C, B and D in Ref.~\cite{Prakash:2016df}, respectively. Below, we verify this claim by using both exact numerical computation of ground states and ITEBD algorithm \cite{Vidal:2007hx}.

\subsection{Exact numerical results}
We exactly compute the ground states of Hamiltonian $H(p,q)$ for systems with up to $N=14$ spins under periodic boundary conditions.
The ground states are obtained by finding the largest eigenvalue of $-H(p,q) + C$ with a sufficiently large constant $C$ and its corresponding eigenvector.
The phase of a state is identified using the following quantities: $A_a \equiv \bra{\psi} u(a)^{\otimes N} \ket{\psi}$,  $A_x \equiv \bra{\psi} u(x)^{\otimes N} \ket{\psi}$, $t\equiv\bra{\psi} \mathcal{T} \ket{\psi}$, $f \equiv \bra{\psi} \mathcal{P}\ket{\psi}$, and the energy gap $\Delta E$ to the first excited state,
where $u(a) = e^{i\frac{2\pi}{3} (S^x+S^y+S^z)/\sqrt{3}}$ and $u(x) = e^{i\pi S^x}$ are generators $a$ and $x$ of the internal symmetry group $A_4$, the operators $\mathcal{T}$ is the translation by one lattice site, and the operator $\mathcal{P}$ is the spatial inversion of spin indices.
When the absolute values of these quantities are unity, the corresponding symmetry is respected by the wavefunciton $\ket{\psi}$.

In order to find the phase diagram, we first consider ground states of a relatively small system size $N=10$ and compute $A_a$, $A_x$, $t$, $p$, and $\Delta E$ for $\sim1000$ different parameters $(p,q)$ randomly spread in the range $p\in [0,2]$ and $q\in[-1/2,1/2]$.
\begin{figure}[tb]
\includegraphics[width=3.5in]{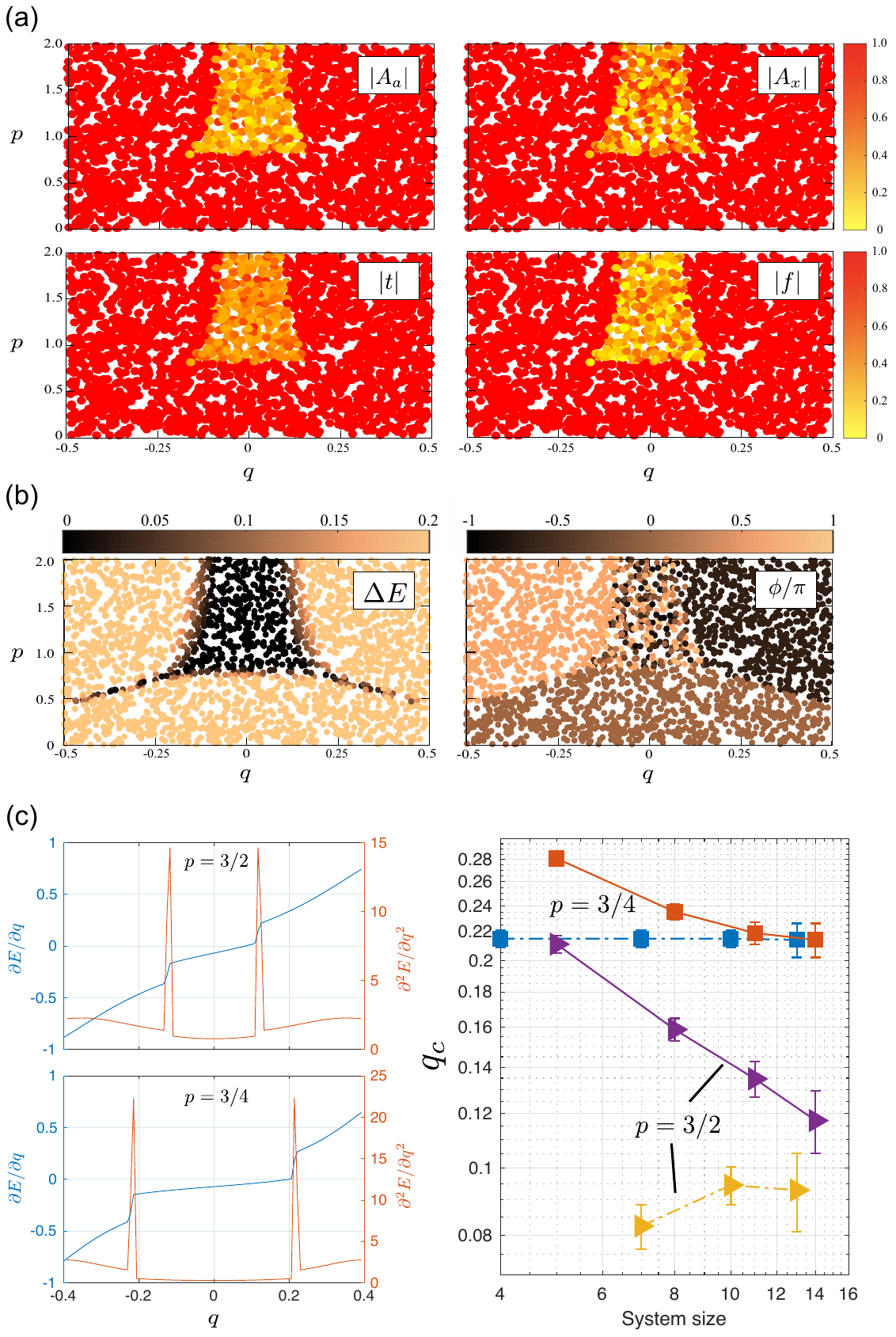}
\caption{Results of exact numerical computations. (a) The absolute values of $|A_a|$, $|A_x|$, $|t|$, and $|f|$ for $p\in [0, 2]$ and $q\in [-1/2,1/2]$ with $N=10$. The ground state respect all symmetries except for near the phase transition points. (b) The energy gap $\Delta E$ and the phase of $|A_a|$ for $p\in [0, 2]$ and $q\in [-1/2,1/2]$ with $N=10$. One clearly finds three separated phases. (c) Extraction of phase transition points using the first and second derivative of ground state energy as a function of $q$ for a fixed value of $p$. Left: the first and second derivatives of energy density $E\equiv E^{(N)}/N$ for a system of $N=14$ spins. Discontinuities in the first derivative or peaks in the second derivative are used as critical points $q_c$. Right: finite size scaling of the extracted critical points. Squares and triangles correspond to extracted critical points for two different cuts $p=3/4$ and $3/2$, respectively. Due to the periodicity of $\chi$ (order 3), the estimated critical points are sensitive to $N\textrm{ mod }3$. For this reason we plot separate scaling curves for different $N\textrm{ mod }3$ and omit $N$ which are integer multiples of $3$.}
\label{sfig:rough_scan}
\end{figure}
The absolute values of $|A_a|$, $|A_x|$, $|t|$, and $|p|$, are shown in Fig.~\ref{sfig:rough_scan}(a), where we find that all symmetries are respected in almost entire range of parameters except $p\gtrsim 1$ and $-0.1\lesssim q|\lesssim 0.1$.
As we will discuss in details below, this domain of  ``symmetry broken'' regime is due to the effects of finite system sizes, rather than being a distinct phase. From closings of energy gaps as well as the complex phase $\phi = \textrm{Im} \left[ \log{(A_a)} \right]$, we clearly identify three distinct phases, I, II, and III  [see Fig.~\ref{sfig:rough_scan}(b)].

In order to extract the phase transition points, we perform finite size scaling analysis.
We obtain the ground state energies $E^{(N)} (p,q)$ of up to $N=14$ spins along two cuts at $p =3/2$ and $3/4$ with $q\in [-0.4,0.4]$.
We compute the first and the second derivatives of energy density $E=E^{(N)}/N$ with respect to the parameter $q$, and extract the phase transition points $q_c$ from discontinuities in $\partial E/\partial q$, or peaks in $\partial^2 E/\partial q^2$.
We note that extracted $q_c$ are sensitive to $N \textrm{ mod } 3$. This is natural since the fiducial SPT phases are distinguished only by a complex phase $\chi \in \{1, e^{\pm i 2\pi/3}\}$ which is periodic in $3$; for a system of $N$ spins, the many-body wavefunction acquires total phase $\chi^N$, which is a function of $N \textrm{ mod } 3$. For example, with $N$ which is an integer multiples of $3$, ground states in three phases are not distinguishable by the complex phase, and we do not expect sharp phase transitions in our  numerics with small system sizes. 

For $p=3/4$, one always finds two phase transitions at $\pm q_c$ which converge to non-zero values in increasing system sizes [Fig.~1(c) right].
In contrast, for $p=3/2$, the critical point $q_c$ decreases with system sizes, suggesting that the phase transition may occur directly from phase II to phase III in thermodynamic limit. 

\subsection{ITEBD}
We further confirm the phase diagram using an independent numerical method based on translationally invariant infinite-size matrix product states (iMPS).
In order to find a ground state, we generally follow ITEBD algorithm introduced in Ref.~\cite{Vidal:2007hx} with a bond dimension $D = 60$. The iMPS $\Gamma$ starts as a random product state. In every iteration, the state is updated after an imaginary time evolution $U_\tau=e^{-\tau  \; H(p,q)}$, where we choose the small time step $\tau = 1/60$. After $300$ repetitions, the updated $\Gamma$ is taken as an approximate ground state. 

\begin{figure}[tb]
\includegraphics[width=6in]{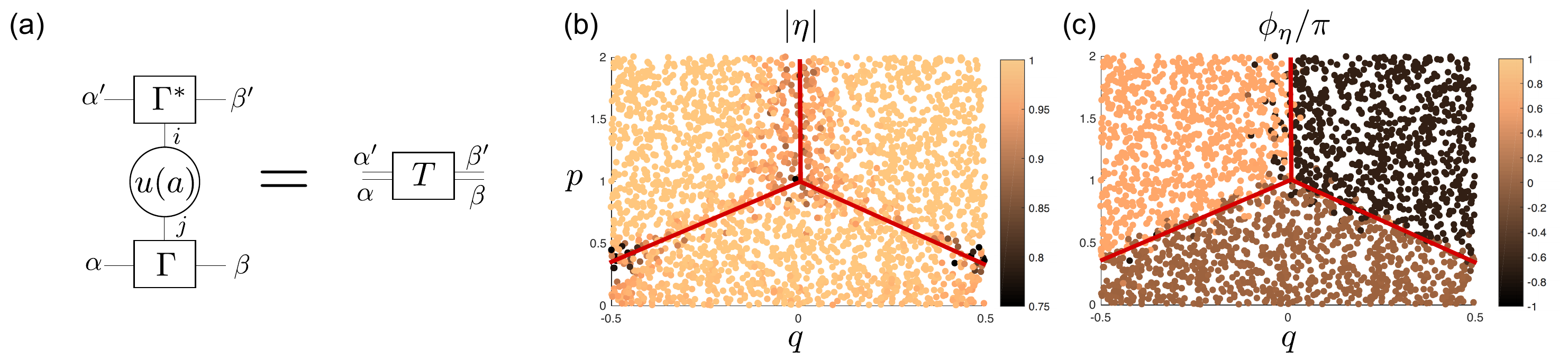}
\caption{ITEBD numerical calculations. (a) Diagrammatic representation of a transfer matrix $T^{\alpha' \beta'}_{\alpha \beta}$. (b-c) Absolute values $|\eta|$ and its phase $\phi_\eta$ as a function of $(p,q)$. The red lines are guides to eyes, for estimated phase boundaries.} 
\label{sfig:ITEBD}
\end{figure}

A SPT phase is identified by studying how  $\Gamma$ transforms under the $120^\circ$ spin rotation $a \in A_4$. Such information is contained the transfer matrix  \cite{Pollmann:2012uj,Prakash:2016df}
\begin{align}
T^{\alpha'\beta'}_{\alpha \beta} = \sum_{ij} u(a)_{ij}\Gamma_{\alpha \beta}^j (\Gamma_{\alpha '\beta'}^i)^*,
\end{align}
where $i, j\in \{\pm1, 0\}$ are physical indices and $\alpha,\alpha',\beta,\beta'\in\{1,\dots D\}$ are virtual (bond) indices [see Fig.~\ref{sfig:ITEBD}(a)].
Since we are interested in an infinite system, we only consider an eigenvalue $\eta$ of the $T^{a'b'}_{ab}$ with the largest absolute value; if $|\eta|=1$ the state $\Gamma$ is invariant under the action of $a$, and otherwise the state breaks the symmetry.
Also, in the case of symmetry unbroken phases, $\eta$ coincides with $\chi(a)$, allowing us to distinguish the three SPT phases.
Figure~\ref{sfig:ITEBD}  (b) and (c) show $|\eta|$ and $\phi_\eta = \textrm{Im}[\log(\eta)]$ for $p\in [0,2]$ and $q\in[-1/2,1/2]$. Clearly, we find three distinct phases separated by red lines.
Near the phase boundaries, $|\eta|$ deviates from 1 probably owing to the finite bond dimension $D$, but there is little evidence for the existence of a symmetry broken phase.

\section{Engineering of $H(p,q)$}
In this section we show that Ising Hamiltonian $H_I = \sum_i S_i^z S_{i+1}^z$ can be engineered into $H(p,q)$ for a range of $(p,q)$ satisfying $2|q|<p<2-2|q|$.
In fact, it is sufficient to show that $H(p,q)$ can be engineered for four points $(p_A,q_A) = (2,0)$, $(p_B,q_B) = (1,-1/2)$, $(p_C,q_C) = (0,0)$, and $(p_D,q_D) = (1,1/2)$ by four pulse sequences. This is because any $H(p,q)$ with $(p,q)$ in the convex hull of those points can be also engineered by concatenating the pulse sequences.

We first start by rewriting the given Ising interactions in a $C$ representation:
\begin{align}
C_I = 
\left(
\begin{array}{cccccccc}
 0 & 0 & 0 & 0 & 0 & 0 & 0 & 0 \\
 0 & 0 & 0 & 0 & 0 & 0 & 0 & 0 \\
 0 & 0 & 0 & 0 & 0 & 0 & 0 & 0 \\
 0 & 0 & 0 & 0 & 0 & 0 & 0 & 0 \\
 0 & 0 & 0 & 0 & 0 & 0 & 0 & 0 \\
 0 & 0 & 0 & 0 & 0 & 0 & 0 & 0 \\
 0 & 0 & 0 & 0 & 0 & 0 & \frac{1}{4} & \frac{\sqrt{3}}{4} \\
 0 & 0 & 0 & 0 & 0 & 0 & \frac{\sqrt{3}}{4} & \frac{3}{4} \\
\end{array}
\right).
\end{align}
Likewise, the $C$-representations of four target interactions are given as
\begin{align}
C_A &=\frac{1}{2}
\left(
\begin{array}{cccccccc}
 1 & -1 & 0 & 0 & 0 & 0 & 0 & 0 \\
 -1 & 1 & 0 & 0 & 0 & 0 & 0 & 0 \\
 0 & 0 & 2 & 0 & 0 & 0 & 0 & 0 \\
 0 & 0 & 0 & 1 & -1 & 0 & 0 & 0 \\
 0 & 0 & 0 & -1 & 1 & 0 & 0 & 0 \\
 0 & 0 & 0 & 0 & 0 & 2 & 0 & 0 \\
 0 & 0 & 0 & 0 & 0 & 0 & \frac{3}{2} & -\frac{\sqrt{3}}{2} \\
 0 & 0 & 0 & 0 & 0 & 0 & -\frac{\sqrt{3}}{2} & \frac{1}{2} \\
\end{array}
\right),
\;\;
C_B= \frac{1}{2}
\left(
\begin{array}{cccccccc}
 1 & 0 & 0 & -1 & 0 & 0 & 0 & 0 \\
 0 & 1 & 0 & 0 & 1 & 0 & 0 & 0 \\
 0 & 0 & 1 & 0 & 0 & 0 & 0 & 0 \\
 -1 & 0 & 0 & 1 & 0 & 0 & 0 & 0 \\
 0 & 1 & 0 & 0 & 1 & 0 & 0 & 0 \\
 0 & 0 & 0 & 0 & 0 & 1 & -\frac{1}{2} & -\frac{\sqrt{3}}{2} \\
 0 & 0 & 0 & 0 & 0 & -\frac{1}{2} & 1 & 0 \\
 0 & 0 & 0 & 0 & 0 & -\frac{\sqrt{3}}{2} & 0 & 1 \\
\end{array}
\right)\\
C_C&= \frac{1}{2}
\left(
\begin{array}{cccccccc}
 1 & 1 & 0 & 0 & 0 & 0 & 0 & 0 \\
 1 & 1 & 0 & 0 & 0 & 0 & 0 & 0 \\
 0 & 0 & 0 & 0 & 0 & 0 & 0 & 0 \\
 0 & 0 & 0 & 1 & 1 & 0 & 0 & 0 \\
 0 & 0 & 0 & 1 & 1 & 0 & 0 & 0 \\
 0 & 0 & 0 & 0 & 0 & 0 & 0 & 0 \\
 0 & 0 & 0 & 0 & 0 & 0 & \frac{1}{2} & \frac{\sqrt{3}}{2} \\
 0 & 0 & 0 & 0 & 0 & 0 & \frac{\sqrt{3}}{2} & \frac{3}{2} \\
\end{array}
\right),
\;\;
C_D = \frac{1}{2}
\left(
\begin{array}{cccccccc}
 1 & 0 & 0 & 1 & 0 & 0 & 0 & 0 \\
 0 & 1 & 0 & 0 & -1 & 0 & 0 & 0 \\
 0 & 0 & 1 & 0 & 0 & 0 & 0 & 0 \\
 1 & 0 & 0 & 1 & 0 & 0 & 0 & 0 \\
 0 & -1 & 0 & 0 & 1 & 0 & 0 & 0 \\
 0 & 0 & 0 & 0 & 0 & 1 & \frac{1}{2} & \frac{\sqrt{3}}{2} \\
 0 & 0 & 0 & 0 & 0 & \frac{1}{2} & 1 & 0 \\
 0 & 0 & 0 & 0 & 0 & \frac{\sqrt{3}}{2} & 0 & 1 \\
\end{array}
\right)
\end{align}
The strengths of isotropic components are given as $s_I=1$,
$s_A=5$, $s_B=4$, $s_C=3$, and $s_D=4$, which fix the rescaling parameters $\beta_a = s_I/ s_a$. As in the case of decoupling spin-1 dipolar interactions, we assume that the set of available unitaries $\mathcal{U}$ is limited to composite pulses made out of up to four $\pm \pi$ and $\pm (\pi/2)$-pulses.  Then, we use a linear programming routing built-in  \verb|Mathematica|. In each case, we find 
a 15-pulse sequence with $\beta^*_A = 1/5$ for $C_A$,
a 12-pulse sequence with $\beta^*_B = 1/4$ for $C_B$,
a 6-pulse sequence with $\beta^*_C = 1/3$ for $C_C$, and 
a 13-pulse sequence with $\beta^*_D = 1/4$ for $C_D$.
These maximum $\beta^*_a$ saturate the required inequalities $\beta^*_a \geq s_I/s_a$ in all four cases. 

\bibliography{refs}